\documentclass[pre,aps,english,twocolumn,notitlepage,superscriptaddress,nofootinbib]{revtex4-2}
\usepackage[T1]{fontenc}  
\usepackage[latin1]{inputenc}

\usepackage{multirow,amsmath}
 
\usepackage{caption}
\usepackage{booktabs}
\usepackage{xcolor}
\usepackage{siunitx}
\usepackage{enumerate}

\usepackage{babel} 
\usepackage[colorlinks,linkcolor =blue,citecolor=blue,urlcolor=blue,bookmarks=false,hypertexnames=true]{hyperref} 
\usepackage{adjustbox}
\usepackage{txfonts}
\usepackage{epsfig,epsf,psfrag} 
\usepackage{graphicx} 
\usepackage{pslatex}
\usepackage{float}
\usepackage{upgreek}
\usepackage{hyperref}
\usepackage{subfigure}
\usepackage{epic,eepic} 
\usepackage{color,pstcol}
\usepackage{pstricks} 
\usepackage{tabularx}
\usepackage{physics}
\usepackage{url}
\usepackage{listings}
\usepackage{color}
\begin{document}

\title{Impurity reveals distinct operational phases in quantum thermodynamic cycles}
\author{Aditya Prakash}
\affiliation{School of Physical Sciences, National Institute of Science Education \& Research, Jatni-752050, India}
\affiliation{Homi Bhabha National Institute, Training School Complex, Anushaktinagar, Mumbai
400094, India}
\author{Abhishek Kumar}\thanks{Current affiliation: Department of Physics, University of Massachusetts, Amherst, MA 01003, USA}
\affiliation{School of Physical Sciences, National Institute of Science Education \& Research, Jatni-752050, India}
\affiliation{Homi Bhabha National Institute, Training School Complex, Anushaktinagar, Mumbai
400094, India}

\author{Colin Benjamin}\email{colin.nano@gmail.com}
\affiliation{School of Physical Sciences, National Institute of Science Education \& Research, Jatni-752050, India}
\affiliation{Homi Bhabha National Institute, Training School Complex, Anushaktinagar, Mumbai
400094, India}
\begin{abstract}
We analyze the effect of impurity on the work output and efficiency of quantum Otto and quantum Carnot heat cycles, modeled as a single quantum particle in an infinite square well (ISW) potential, which is the working substance. We solve this quantum mechanical system perturbatively up to first and second order in strength of the impurity for strong and weak coupling regimes, respectively. We derive the analytical expressions of work and efficiency for the strong coupling regime to the first order in the strength parameter. The threshold value of the strength parameter in weak coupling is obtained up to which the numerical result agrees with the perturbative result for a repulsive and attractive impurity. To our surprise, an embedded impurity unlocks new operational phases in the system, such as a quantum heat engine, quantum refrigerator, and quantum cold pump. In addition, the efficiency of the quantum Otto heat engine is seen to reach Carnot efficiency for some parameter regimes. The cooling power and coefficient of performance of the quantum refrigerator and quantum cold pump are non-trivially affected by the impurity.
\end{abstract}
\maketitle

\section{Introduction}
A classical heat engine performs work via classical thermodynamic processes, while a quantum heat engine (QHE) does the same using quantum thermodynamic processes. There has been a long history of research in quantum thermodynamic processes, as shown in Refs.~\cite{bender2000quantum,bender2002entropy,quan2007quantum,quan2009quantum}. QHE's such as quantum Carnot heat engine (QCHE), quantum Otto heat engine (QOHE), and quantum Stirling heat engines (QSHE) have been studied with different working substances, e.g., a particle in an infinite square well (ISW) potential~\cite{quan2007quantum}, a particle in the harmonic oscillator potential~\cite{quan2007quantum}, spin systems~\cite{PhysRevLett.100.140501,huang2014quantum,das2019measurement}, stanene~\cite{fadaie2018topological}, strained-graphene~\cite{PhysRevE.96.032118}, Dirac particles~\cite{PhysRevE.86.061108,PhysRevE.94.022109}, two-level system~\cite{kieu2006quantum}, multi-level system~\cite{quan2005quantum}, a continuum working medium ~\cite{li2006quantum}, photon gas~\cite{hardal2015superradiant} etc. Different working substances are used, and different heat baths are used in Refs.~\cite{niedenzu2018quantum, de2019efficiency} for the QOHE. Ref.~\cite{niedenzu2018quantum} establishes an efficiency bound for the QOHE, which surpasses the Carnot efficiency bound. In another study Ref.~\cite{de2019efficiency} it is shown that a QOHE operating between an effective negative temperature and effective positive temperature is more efficient than when both are at positive temperatures. In all these studies, a missing element has been the effect of impurity on work output and efficiency of the quantum heat engines.

Recent works on the QSHE Ref.~\cite{PhysRevLett.106.070401, kim2012szilard, thomas2019quantum} use an impurity in the strong coupling limit for the insertion and removal of a barrier to separate a box into two parts and then merge the two compartments into one. It was observed in Ref.~\cite{ thomas2019quantum} that Carnot efficiency is approached in the low-temperature case by insertion and removal of a barrier during the thermodynamic cycles of QSHE. Work output and efficiency are calculated by varying multiple parameters of the box, including the temperature of baths, length of the box, the strength of the impurity, and position of the impurity. The expectation is impurities will reduce the power output, leading to more scattering. However, in our study of the effect of an impurity in quantum Otto and quantum Carnot thermodynamic cycles, we find that the work output and efficiency are enhanced in specific parameter regimes of strength and position of the impurity. Regions with negative work~\cite{das2019measurement} are obtained, which are interpreted as quantum refrigerators and quantum cold pumps depending on the sign of the heat exchanged with reservoirs.

To summarize the main findings of this paper, the perturbative results for energies of the infinite square well (ISW) up to second order for weak coupling and up to first order for strong coupling strength parameters of the impurity are obtained. Next, the agreement between the numerical and the perturbative spectrum is established for attractive and repulsive impurities. Then, the work output and efficiency plots were obtained over a range of strength, temperature and length values in the weak and strong coupling regimes. We notice an enhancement in work and efficiency due to the impurity for both QOHE and QCHE. Based on the energy flow direction, the negative work regions for both the cycles act as cold pumps or refrigerators. We concentrate our study on Otto and Carnot cycles, as they are widely used engines in the classical world. Otto cycle is commonly used in automobile engines. The outline of this paper is as follows. We begin by solving the Hamiltonian for ISW with impurity by providing an exact solution for the energy eigenvalues via a perturbative correction to first and second-order in strong and weak coupling regimes, respectively. We do compare the weak coupling perturbative result with the numerical result obtained after solving the transcendental dispersion relation for verification. In section \ref{results}, we show the detailed calculation of the work output and efficiency for both the quantum Otto cycle (QOC) and quantum Carnot cycle (QCC). We find out what happens to work output and efficiency as we vary various parameters of the system, including the strength of impurity, length of ISW, the temperature of the bath, and position of the impurity? In section \ref{analysis} we analyze the work results via tables which bring out the effect of impurity. We end with a conclusion, highlighting the impact of impurity and some possible experimental realization. Following are some acronyms that we will use throughout the paper: infinite square well (ISW), quantum heat engine (QHE), quantum Otto cold pump (QOCP), quantum Otto heat engine (QOHE), quantum Otto refrigerator (QOR), quantum Carnot heat engine (QCHE), quantum Carnot cold pump (QCCP), quantum Carnot refrigerator (QCR), quantum Otto cycle (QOC), quantum Carnot cycle (QCC).
\section{Theory}
\subsection{The model Hamiltonian and solution}
\begin{figure}[H]
\begin{center}
\includegraphics[width=5cm, height = 5cm]{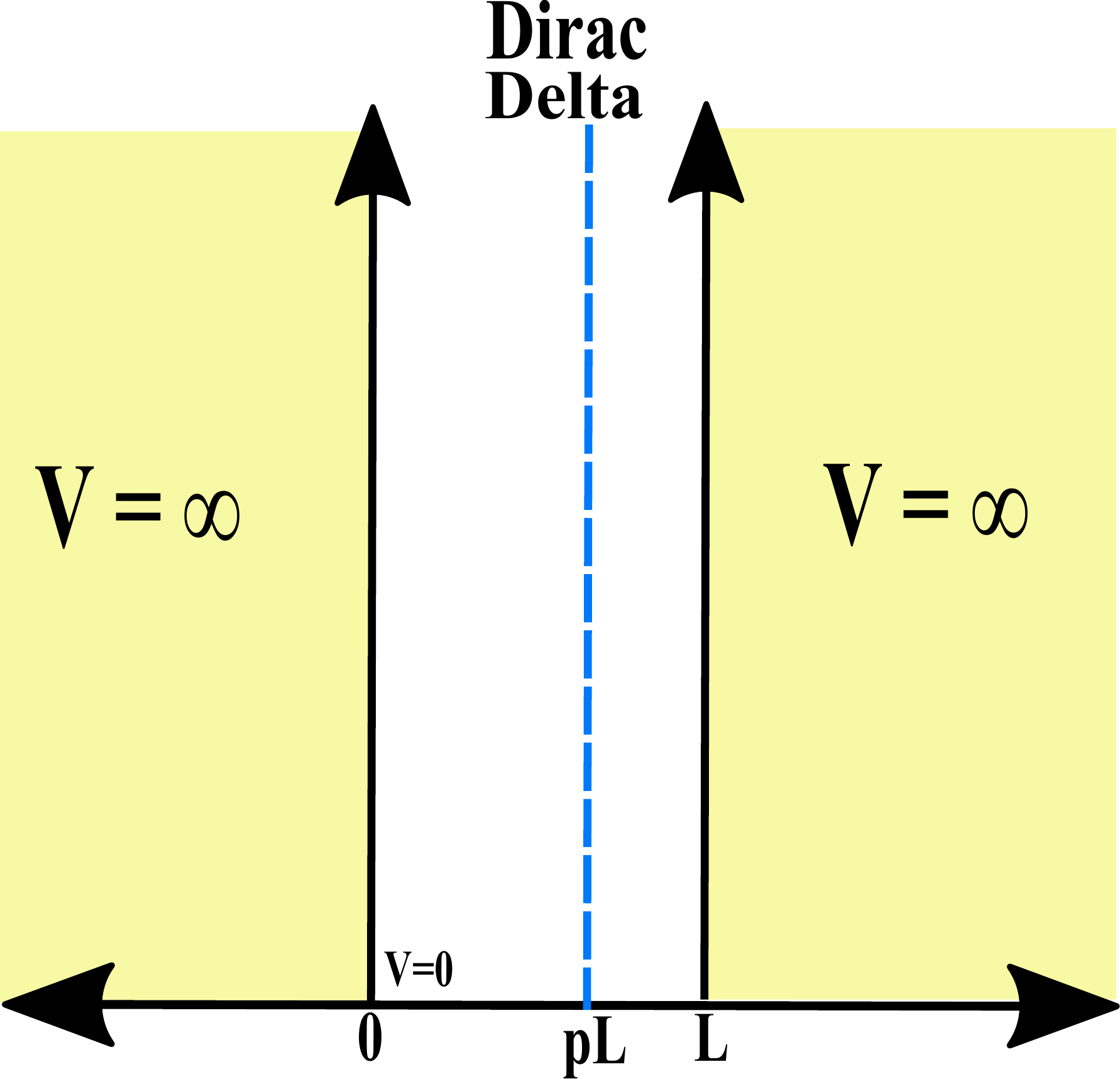}
\end{center}
\caption{The potential $V(x)$ of ISW embedded with impurity at position $pL$ versus $x$. The length of the well is $L$ and strength $\lambda$ of the impurity.}
\label{fig:del}
\end{figure}

The Hamiltonian with a particle of mass $m$ in a 1-D infinite square well (ISW) of length $L$ and an impurity at position $pL$ ($0\leq p\leq1$) inside the well is given by,
\begin{eqnarray}
H &=& H_{0}+H^{\prime},\nonumber\\
\mbox{with, } H_{0} &=& -\frac{\hbar^{2}}{2m}\nabla^{2} + V(x),\mbox{ }
V(x)= \begin{cases}
0 &\text{for } 0\leq x \leq L\\
\infty &\text{otherwise}
\end{cases},\nonumber\\
\mbox { and }
H^{\prime} &=& -\lambda \delta(x - pL).
\label{H}
\end{eqnarray}
$H^{\prime}$ denotes impurity modeled as a $\delta-$function potential, with $p$ determining the position of the impurity inside the well and $\lambda$ represents the strength of the impurity. $\lambda < 0$ (> 0) implies repulsive (attractive) delta function which behaves as a barrier (well). Figure \ref{fig:del} shows the ISW potential with a repulsive impurity.

\subsection{The exact solution}
The aim here is to find the solution of the time-independent Schr\"{o}dinger equation $H\Psi = E\Psi$, with the wave function $\Psi$ satisfying the following three boundary conditions:
\begin{eqnarray}
\lim_{x\to pL-}\Psi(x)& =& \lim_{x\to pL+}\Psi(x),
\label{ac}\\
\lim_{x \to pL+}\frac{d\Psi(x)}{dx} &-& \lim_{x \to pL-}\frac{d\Psi(x)}{dx} = -\frac{2m\lambda}{\hbar^{2}} \Psi(pL),
\label{ab}\\
\Psi(0)&=&\Psi(L)=0.
\label{bc}
\end{eqnarray}
For positive energies $E=\frac{\hbar^2 k^2}{2m} > 0 $, the wave function solution of the Schr\"{o}dinger equation satisfying the boundary condition in Eqs.~(\ref{ac},\ref{bc}) has the form-
\begin{equation}
\Psi(x) = \begin{cases}
A\sin(k(L-x))\sin(kpL), &\text{for }pL \leq x \leq L,\\
A\sin(kx)\sin(kL(1-p)), &\text{for }0 \leq x < pL,\\
0 &\text{otherwise}.
\end{cases}
\label{wave1}
\end{equation}
where $A$ denotes normalization constant for wavefunction and $k = \sqrt{\frac{2mE}{\hbar^2}}$.\\
Applying Eq.~(\ref{ab}) and using the wave function obtained in Eq.~(\ref{wave1}) gives us the dispersion relation, see also Ref.~\cite{smith2010effect} as-
\begin{equation}
(kL)f\sin (kL) = 2\sin (kpL)\sin (kL(1-p)),
\label{dis+}
\end{equation}
where $f = \frac{\hbar^{2}}{m\lambda L}$ is a dimensionless parameter which is a measure of the strength of the impurity.

\begin{figure}
\centering
\includegraphics[height =8cm , width = 8cm]{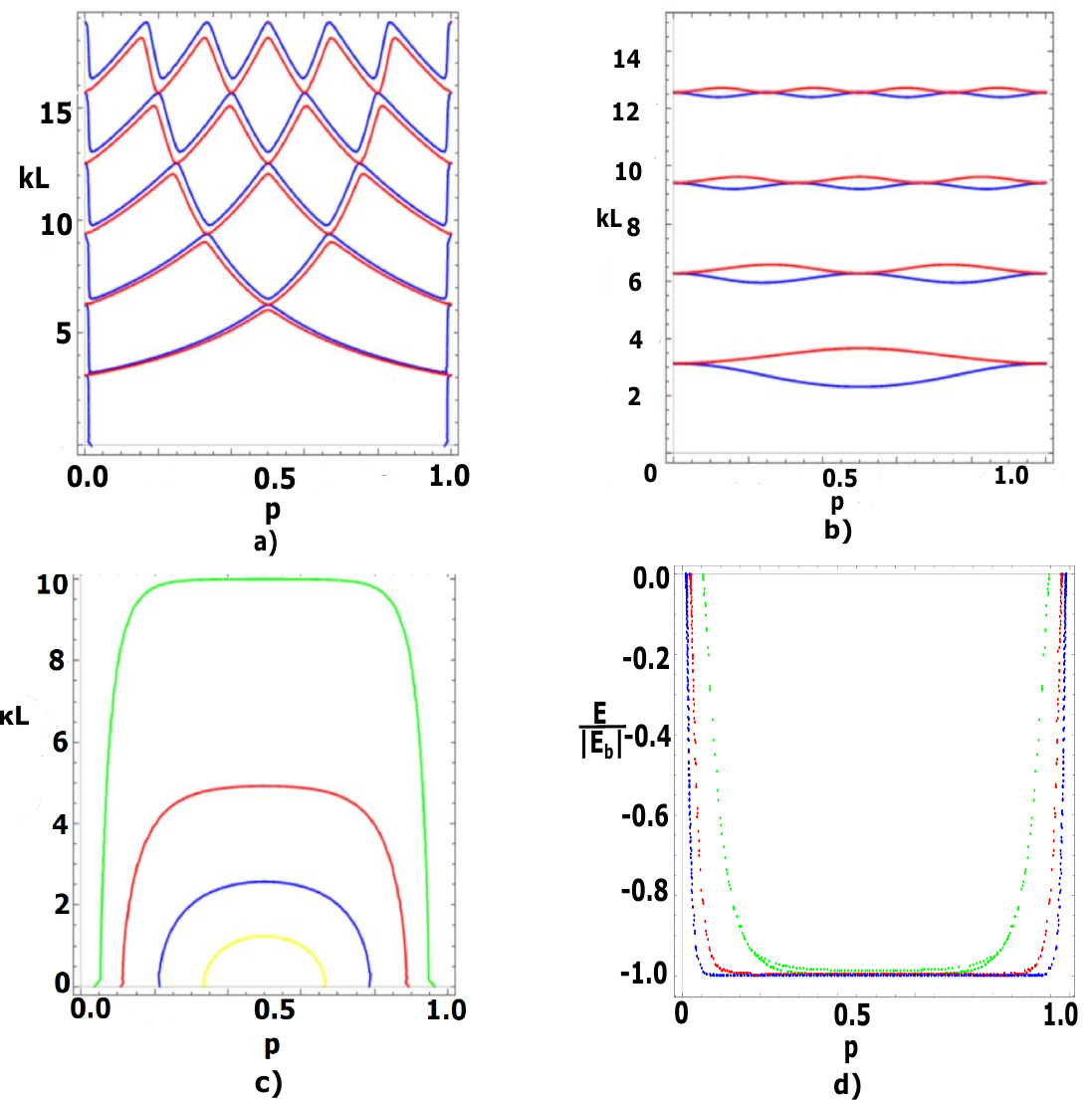}
\caption{\textbf{a)} $kL$ vs position of the impurity $p$ (for strong coupling case), plot for attractive impurity $f = 0.02$ (in blue) and plot for repulsive impurity $f = -0.02$ (in red). \textbf{b)} $kL$ vs $p$ (for weak coupling case) plot for attractive impurity $f = 1$ (in blue) and repulsive impurity $f = -1$ (in red).\textbf{c)} $\kappa L$ vs $p$ for four different strength parameters $f = 0.44$, $0.33$, $0.2$, and $0.1$ (yellow, blue, red, and green). \textbf{d)} $\frac{E}{|E_{b}|}$ vs $p$ at three different strength parameters $f = 0.1,$ $0.04,$ and $0.02$ (green, red, and blue). }
\label{fig2}
\end{figure}

The energy spectrum resulting from the dispersion in Eq.~(\ref{dis+}) are plotted in Figs.~\ref{fig2}(a),~\ref{fig2}(b). These figures are plotted for different values of the strength parameter ($f$), and each figure contains both repulsive and attractive spectrum with the same magnitude of the strength parameter. Fig.~\ref{fig2}(a) sets the parameter $f$ near the strong coupling limit ($|f| \ll 1 $) while Fig.~\ref{fig2}(b) is set in the weak coupling regime ($|f| > 0.5$). For both the cases, we see that the separated repulsive and attractive spectrum in Figs.~\ref{fig2}(a),~\ref{fig2}(b) become degenerate when the strength is close to the extreme strong coupling limit ($|f| \ll 0.1$) or weak coupling limit ($|f| \geq 0.5$). However, for an attractive impurity in a strong coupling case, we have negative energy, which is not observed in any other case. In the next section, we will discuss the strong and weak coupling limits in more detail.

Now for negative energies we have $\kappa=\sqrt{-\frac{2mE}{\hbar^2}}$ with E<0, and carrying out a similar procedure as above we obtain-
\begin{equation}
\Psi(x) = \begin{cases}
A\sinh(\kappa(L-x))\sinh(\kappa pL), &\text{for }pL \leq x \leq L,\\
A\sinh(\kappa x)\sinh(\kappa L(1-p)), &\text{for }0 \leq x < pL,\\
0 &\text{otherwise}.
\end{cases}
\end{equation}
and the dispersion relation, see also Ref.~\cite{smith2010effect},as
\begin{equation}
(\kappa L)f\sinh(\kappa L) = 2\sinh(\kappa pL)\sinh(\kappa L(1-p)).
\label{dis-}
\end{equation}

For the case of the attractive delta potential ($\lambda > 0$) the solutions may have both positive and negative energies while for the repulsive delta potential ($\lambda < 0$), the solutions will only have positive energies since $E > V_{min}$ (Ref.~\cite{griffiths2005introduction}) is the requirement to obtain a normalizable solution to the time-independent Schr\"{o}dinger equation where $V_{min}$ is minimum of the potential ``$H^{\prime} + V(x)$''. For the repulsive case $V_{min} = 0$ and hence there will be no negative energy for all values of the strength parameter $f$. The dispersion relations in Eqs.~(\ref{dis+},\ref{dis-}) are transcendental equations and can be solved numerically, see Ref.~\cite{joglekar2009particle,smith2010effect}.
We find that the numerical solutions to the dispersion relation for negative energies in Eq. (\ref{dis-}) exist only for attractive impurity. Further, from Fig.~\ref{fig2}(c) we observe that negative energy solutions exist for attractive impurity only when $f<|0.5|$. There are no solutions to Eq. (\ref{dis-}) for $f\geq0.5$ even for attractive impurity. Unlike the attractive delta potential without ISW potential where negative energy bound state ($E_{b} = -\frac{\hbar^{2}}{2mL^{2}}\frac{1}{f^{2}}$) exists at all strengths, here we have negative energy solution possible only in a certain range of strength parameter, i.e., $|f| \leq 0.5$ and this can be seen by looking at the trend in Fig.~\ref{fig2}(c).
In Fig.~\ref{fig2}(d) energy becomes flat at $E_{g} = - E_{b}$ and maximum flatness is obtained at lower values of $f$. In the extreme cases of strong coupling where $f\approx 0$, one can use $E = -E_{b}$ independent of position of the impurity.

One also realizes that when impurity is near the walls, see also Ref.~\cite{smith2010effect}, the eigenvalues should reduce to just ISW eigenvalues $\left(E_{n} = \frac{n^{2}\pi^{2}\hbar^{2}}{2mL^{2}}\right)$. This situation can physically be thought of as the impurity getting merged with the wall of the ISW. From the dispersion relations in Eqs.~(\ref{dis+},\ref{dis-}) it can be clearly seen that $E(k,p) = E(k,1-p)$ and $E(\kappa,p) = E(\kappa, 1-p)$. Thus, the spectrum is symmetric about $p=1/2$ and $\frac{dE}{dp}\big\rvert_{p=1/2} = 0$. In the next subsection, we will look at the perturbative analytical solution for different strengths and check the agreement with numerical results.

\subsubsection{The perturbative solution for weak coupling regime ($|\lambda| \ll 1$ or $|f|\geq0.5$ )}
In case of weak coupling ($|\lambda| \ll 1$), the perturbative eigenenergy correction up to second order cam be derived by using expansion of $k$ ($= k_{0} + \lambda k_{1} + \lambda^{2} k_{2} + \dots $) and $\kappa$ ($=\kappa_{0} + \lambda\kappa_{1} + \lambda^{2}\kappa_{2} + \dots$). This is not the same as Rayleigh-Schr\"{o}dinger perturbation theory expansion of eigenenergy ($E_{n} = E_{n}^{(0)} + \lambda E_{n}^{(1)} + \lambda^{2}E_{n}^{(2)} + \dots$) where we obtain recursive formula for eigenenergy corrections. However, both the methods will lead to the same result. It is difficult to find closed form formula for the second order eigenenergy correction using Rayleigh-Schr\"{o}dinger perturbation theory since it involves restricted summation over large number of states. But we can easily get a closed form formula by $k$ (and $\kappa$) expansion method, as shown in Ref.~\cite{bera2008perturbative}. Applying the $k$ expansion to the dispersion relation in Eq.~(\ref{dis+}), we have,
\begin{eqnarray}
\alpha\sin (\alpha L)=\frac{2m\lambda}{\hbar^{2}} \sin (\alpha pL)\sin(\alpha (1-p)L)
\label{expand}
\end{eqnarray}
where $\alpha = k_{0}+\lambda k_{1} + \lambda^{2} k_{2} + \dots$\\
Using the power series expansion of $\sin$ function in Eq.~(\ref{expand}) and collecting powers of $\lambda^{0},$ $\lambda$, and $\lambda^{2}$, we get,
\begin{eqnarray}
&k_{0}&\sin (k_{0} L) = 0 \label{lam0},\\
k_{1} &=& -\frac{2m}{\hbar^{2}}\frac{\sin^{2} (k_{0}pL)}{k_{0}L}\label{lam1}, and \\
k_{2} = -\frac{k_{1}^{2}}{k_{0}} &+& \frac{2m}{\hbar^{2}}\frac{k_{1}}{k_{0}}(1-2p)\sin(k_{0}pL)\cos(k_{0}pL).
\label{lam2}
\end{eqnarray}

We know, $k_{0} \neq 0$, otherwise zero order solution will vanish. Thus, $\sin (k_{0}L) = 0$ which implies $k_{0} = \frac{n\pi}{L}$, ($n =$ 1, 2, $\dots$) , this relation is further used in Eqs.~(\ref{lam1}),(\ref{lam2}). The first and second order eigenenergy correction is obtained from
\begin{equation}
E_{n} = \frac{\hbar^{2}(k_{0} + \lambda k_{1} + \lambda^{2} k_{2} + \dots)^{2}}{2m},
\label{eigen}
\end{equation}
with $E_{n}^{(1)}$ and $E_{n}^{(2)}$ given by,
\begin{equation}
E_{n}^{(1)} = \frac{\hbar^{2}k_{0}k_{1}}{m}, E_{n}^{(2)} = \frac{\hbar^{2}}{2m}[k_{1}^{2} + 2k_{0}k_{2}]\label{eigen2}.
\end{equation}
Substituting $k_{0}$, $k_{1}$, and $k_{2}$ from Eqs.~(\ref{lam0}, \ref{lam1} and \ref{lam2}) in Eq.(~\ref{eigen}), we can write eigenenergy up to second order as,
\begin{eqnarray}
&&E_{n}(p,f) = \frac{\hbar^{2}}{2mL^{2}}\left[(n\pi)^{2} - \frac{4\sin^{2}(n\pi p)}{f}\right] \nonumber\\ &-&\frac{\hbar^{2}}{2mL^{2}}\left[\frac{4\sin^{4} (n\pi p)}{n^{2}\pi^{2}f^{2}}(1 + 2n\pi(1-2p)\cot(n\pi p))\right]
\label{eigenenergy}
\end{eqnarray}
This is the eigenenergy expression for weak coupling attractive ($f > 0$) and repulsive ($f < 0$) impurity delta function. Eq.~(\ref{eigenenergy}) in the limit $f \to \infty$ (i.e. vanishing impurity), reduces to ISW potential eigenenergy solution. The exact same eigenenergy expression is obtained for the expansion using $\kappa$, this can be easily verified by replacing $k$ with $i\kappa$ in the above calculation. An eigenenergy expression in the strong coupling case where $|\lambda| \gg 1$, can also be obtained using an expansion of the form $k = k_{0} + \frac{1}{\lambda}k_{1} + \frac{1}{\lambda^{2}}k_{2} +\dots$ and the details of the calculation for strong coupling eigenenergy results are presented in the next subsection.

The first order pertubative results in Eq.({\ref{eigenenergy}}) can be verified by applying Rayleigh-Schr\"{o}dinger time-independent perturbation theory. The first order energy correction is found by treating $H^\prime$ as a perturbation in Eq.~(\ref{H}). The eigenfunctions and eigenvalues of the infinite potential well in absence of an impurity Ref.~\cite{griffiths2005introduction} are easy to get-
\begin{equation}\Psi_{n}(x) = \sqrt{\frac{2}{L}}\sin\left(\frac{n\pi x}{L}\right),\text{ } E_{n} = \frac{n^2\pi^2\hbar^2}{2mL^2}, n=1,2,3...\label{En}\end{equation}

The first order correction, for both attractive and repulsive $\delta$ function impurity to the energy eigenvalues can be calculated easily Ref.~\cite{sakurai1995modern} and they are,
\begin{equation}
\begin{array}{l}
E_{n}^{(1)} = \langle\Psi_{n}|H^{\prime}|\Psi_{n}\rangle = -\frac{2}{L}\int_{-\infty}^{\infty}\sin^2\left(\frac{n\pi x}{L}\right)\delta(x-pL) dx\\
\implies E_{n}^{(1)} = -\frac{2}{L}\sin^2(n\pi p).
\end{array}
\end{equation}

This is the same first order correction as was obtained in Eq.~(\ref{eigen2}) using the expansion method. Hence, the energy levels $E_{n}$ of the ISW potential with an attractive ($f > 0$) or repulsive ($f < 0$) $\delta$-function impurity up to first order are:

\begin{equation}
E_{n} = \frac{\hbar^2}{2mL^2}\left(n^{2}\pi^{2} - \frac{4\sin^{2} (n\pi p)}{f}\right), \text{ where $n$ = 1,2,3,...}
\label{perturb}
\end{equation}

\begin{figure}
\includegraphics[width = 8cm ,height = 5cm]{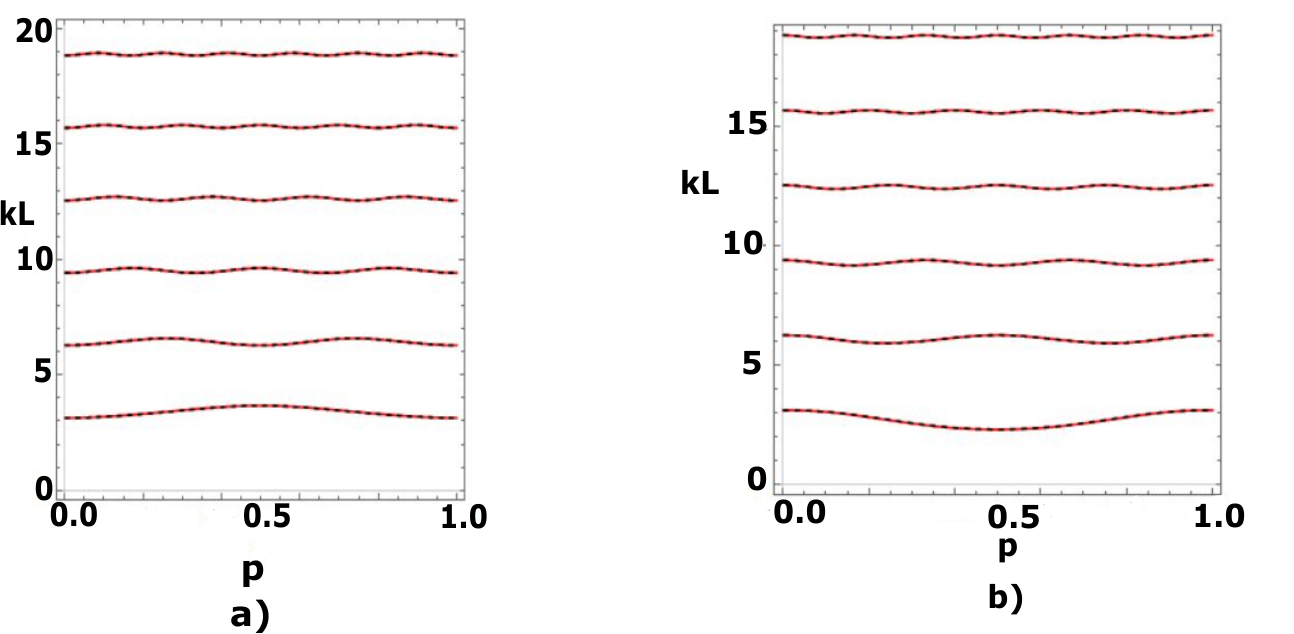}
\caption{Weak coupling pertubative results. \textbf{a)} $kL$ vs $p$ for the attractive $\delta$ function impurity with strength $f = 1$ . The first six energy levels are plotted both numerically (red solid) and perturbatively (black dashed). \textbf{b)} $kL$ vs $p$ for the repulsive $\delta$ function with the strength parameter $f = -1$.}
\label{Figure3}
\end{figure}

Eq.~(\ref{perturb}) is the same as Eq.~(\ref{eigenenergy}) if we restrict to first order. The perturbative relation of Eq.~(\ref{eigenenergy}) is compared with the numerical results of Eq.~(\ref{dis+}) in Fig~\ref{Figure3}. We find both the equations are in good agreement when $|f|\geq 0.5$. Fig. 3(c) shows the ratio of ground state energy to the bound state energy ($E_{b} = -\frac{\hbar^{2}}{2mL^{2}}\frac{1}{f^{2}}$) plotted varying position $p$ of the impurity for both attractive and repulsive impuirty strength, and it can be seen that the numerical curve for attractive and repulsive $\delta$ function potential completely overlaps with the perturbative curve for strength $|f| \geq 0.5$. But at values of strength parameter $|f| < 0.5$, the numerical result for the attractive and repulsive $\delta$ function deviates from the analytical perturbative result. Figs. 3(a) and 3(b) represent plots of the first six energy levels as function of position $p$ of the impurity for attractive (repulsive) $\delta$ function potential both numerically and perturbatively where the strength parameter $|f| \geq 0.5$. We see that the two results completely overlap with each other, and hence we are free to use the perturbative results if $|f| \geq 0.5$. The weak coupling perturbative results are used in the next section to calculate work output of QOC and QCC. We notice that the eigenenergy relation in Eq.~(\ref{eigenenergy}) is symmetric about $p = 1/2$ as it satisfies, $E_{n}(f,p) = E_{n}(f, 1-p)$ and when $p$ is close to zero the eigenvalues reduce to ISW potential eigenvalue solution.

\begin{figure}
\includegraphics[width = 7cm ,height = 6cm]{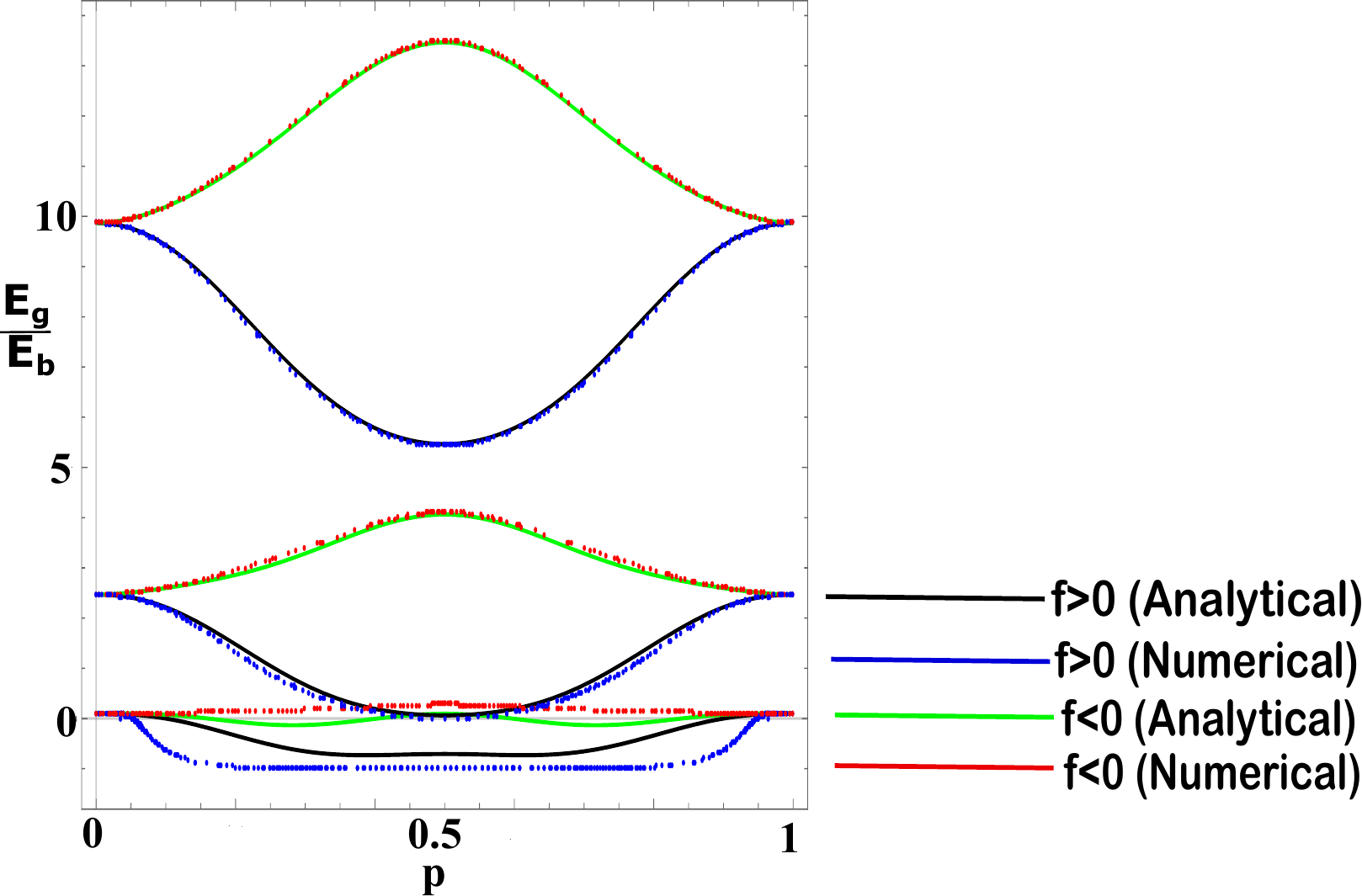}
\caption{$\frac{E_{g}}{|E_{b}|}$ vs $p$ where black ($f > 0$) and green solid lines ($f < 0$) are the perturbative ground state results for the attractive and repulsive $\delta$ function respectively which is plotted as a function of position $p$ of impurity. There are four colours green, red, blue and black. Green ($f<0$) and black ($f>0$) are plotted analytically using perturbed ground state expression, whereas red ($f<0$) and blue ($f>0$) are plotted numerically. The lowest, middle and the topmost blue and red dots pair or black and green solid line pair are plotted at the strength parameter $|f|$ = 0.1, 0.5 and 1 respectively. }
\label{Figure4}
\end{figure}

In Fig.~\ref{Figure4} the lowest dotted pair does not match the solid line pair, but the other two dotted pairs overlap with the solid line pairs. We can conclude using this plot that the ground state perturbative results agree with the numerical results for the strength parameter $|f| \geq 0.5$. Thus, all the weak coupling perturbative calculations will be performed in the regime $|f| \geq 0.5$.\\

\subsubsection{The perturbative solution for strong coupling regime ($|\lambda| \gg 1$ or $|f|\ll0.1$ )}
Impurity modeled as a barrier in a box was solved in Ref.~\cite{thomas2019quantum} for a strong coupling case ($f \to 0$) but with its position fixed at the center of the ISW. In this paper, we generalize the impurity position to any arbitrary place in the ISW. We derive the energy eigenvalue for the strong coupling case by applying perturbation up to the first order in the strength parameter.

In extreme strong coupling case ($|\lambda| \gg 1$ or $|f| \ll 0.1$ ) the expansion of $k = k_{0} + \frac{1}{\lambda}k_{1} + \frac{1}{\lambda^{2}}k_{2} + \dots$ is applied to the dispersion relation (in Eq.~\ref{dis+}), and we obtain:
\begin{eqnarray}
\beta \sin (\beta L) = \frac{2m\lambda}{\hbar^{2}} \sin (\beta pL)\times& \sin(\beta(1-p)L)
\label{expand1}
\end{eqnarray}
where $\beta = k_{0}+\frac{1}{\lambda} k_{1} + \frac{1}{\lambda^{2}} k_{2} + \dots$
Calculating the power series expansion of $\sin$ and collecting powers of $\lambda^{1}$ on both sides,
\begin{equation}
\sin (k_{0}pL)\sin (k_{0}(1-p)L) = 0,
\label{strong1}
\end{equation}
which implies either $\sin (k_{0}pL) = 0$ or $\sin (k_{0}(1-p)L) = 0$. Choosing $\sin(k_{0}pL) = 0$, gives ($ k_{0} = \frac{n\pi}{pL}$, where n = 1,2,3,$\dots$, then collecting powers of $\lambda^{0}$, we have,
\begin{equation}
(k_{0} - \frac{2m}{\hbar^{2}}k_{1}pL)\sin (k_{0}L) = 0.\label{strong2}\\
\end{equation}
If $\sin (k_{0}L) \neq 0$, then $k_{1} = \frac{\hbar^{2}k_{0}}{2mpL}$
Thus the eigenenergy correction up to first order is given by,
\begin{equation}
E_{n} = \frac{\hbar^{2}(k_{0} + \frac{1}{\lambda} k_{1} + \dots)^{2}}{2m},\end{equation}
\begin{equation}
\text{or},
E_{n} = \frac{\hbar^{2}}{2mL^{2}}\left((k_{0}L)^{2} +\frac{2}{\lambda}(k_{0}L)(k_{1}L)+ \dots,\right)
\label{strong}
\end{equation}
substituting $k_{0}$ and $k_{1}$ in Eq.~(\ref{strong}), we get the strong coupling eigenenergy upto first order in strength parameter ($f$) as,
\begin{equation}
E_{n}(f, p) = \frac{\hbar^{2}}{2mL^{2}}\left(\frac{n^{2}\pi^{2}}{p^{2}} + \frac{nf\pi}{p^{3}}\right).
\label{strongeigen}
\end{equation}
Choosing $\sin (k_{0}(1 - p)L) = 0$ instead of $\sin (k_{0}pL) = 0$, means one needs to replace $p$ by $1 - p$ in Eq.~(\ref{strongeigen}). In that case, the strong coupling eigenenergy is $E_{n}(f, 1-p)$.

\subsection{Quantum Thermodynamics}
A combination of different quantum thermodynamic processes and number of strokes in each cycle results in different types of quantum heat engines, see Refs.~\cite{quan2007quantum,quan2009quantum}. Before starting the calculation of work output and efficiency, let us discuss the thermodynamics of these. The internal energy, see Ref.~\cite{quan2007quantum}, depends on the temperature in the case of classical ideal gas and the number of degrees of freedom, but for the quantum mechanical system it depends on other parameters. Total energy is $U = Tr[\rho H]$ where for distinguishable particles, density matrix is $\rho = \sum_{n} P_{n}|n\rangle\langle n| = \sum_{n} \frac{\exp (-\beta E_{n})}{Z}|n\rangle\langle n|$ and $Z = Tr(\exp (-\beta H))$ is the partition function. Thus $U = \sum_{n}P_{n}E_{n}$, where $P_{n}$ is the occupation probability of the $n^{th}$ eigenstate and $E_{n}$ is the $n^{th}$ eigenenergy of the working substance. We have $dU = \sum_{n}(E_{n}dP_{n} + P_{n}dE_{n})$. The first law of thermodynamics is $dU = \textit{\dj}Q + \textit{\dj}W$,
where $U$ is a state function and $Q$, $W$ are path-dependent functions and this has caused the notational change in the differentials. Thus,
\begin{equation}
\textit{\dj}Q = \sum_{n}E_{n}dP_{n} \text{~~~~ and ~~~~} \textit{\dj}W = \sum_{n}P_{n}dE_{n}.
\label{prime}
\end{equation}
At thermal equilibrium, we can write $\textit{\dj}Q = T dS$. Eq.~(\ref{prime}) holds true for both equilibrium and non equilibrium case.
In the rest of the paper, we will compute the work output and efficiency of QOC and QCC (when the reversibility condition is satisfied for the Carnot cycle). Our working substance would be ISW with a Dirac delta impurity.
The particle mass $m$ is equal to the electron mass. Temperatures of hot and cold reservoirs are denoted by $T_{h}$ and $T_{c}$.

\subsection{Categorization of the System}
Depending on the signs of $Q_{in}$, $Q_{out}$ and work done, the cycle can be categorized into heat engine, refrigerator, Joule pump, cold pump as also shown in Ref.~\cite{josephson}.

\begin{table}
\scalebox{0.88}{
\begin{tabular}{|l|l|l|l|}
\hline
 \includegraphics[width=2.3cm, height = 3cm]{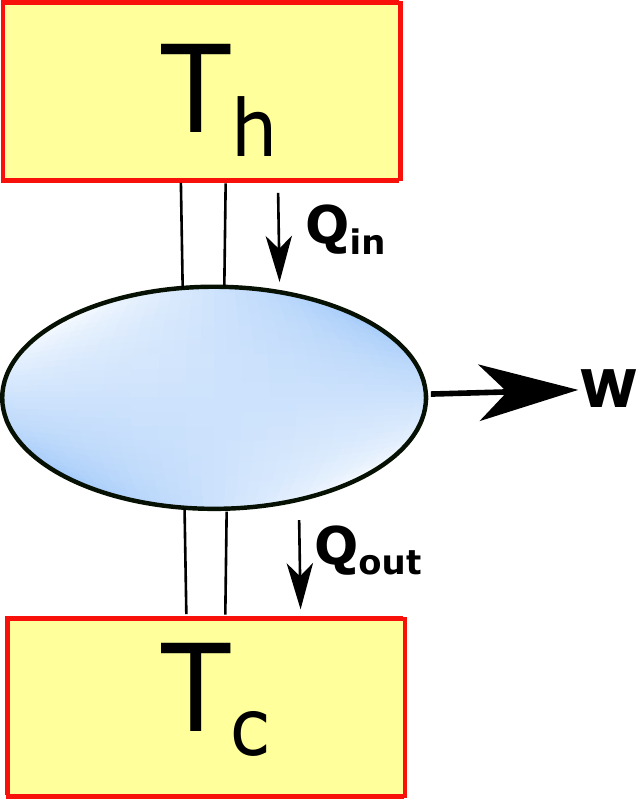} &  \includegraphics[width=2.3cm, height = 3cm]{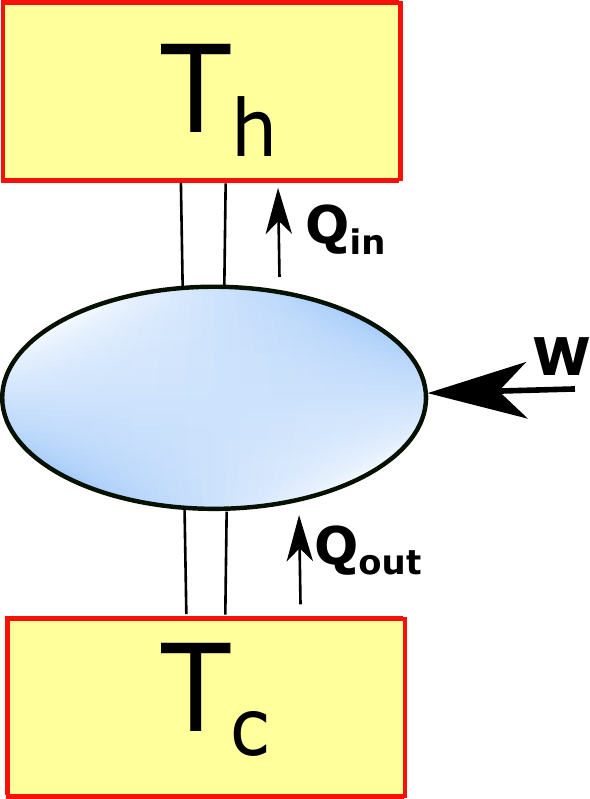}  & \includegraphics[width=2.3cm, height = 3cm]{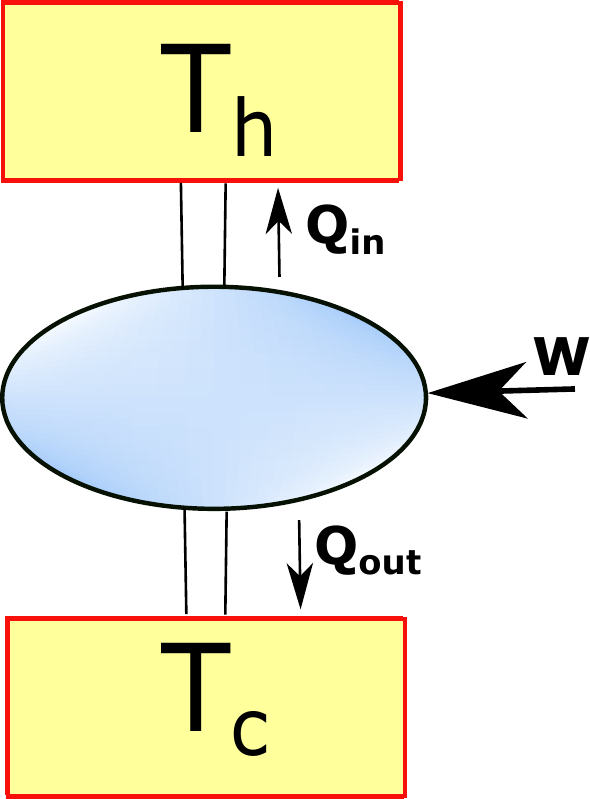}  & 
 \includegraphics[width=2.3cm, height = 3cm]{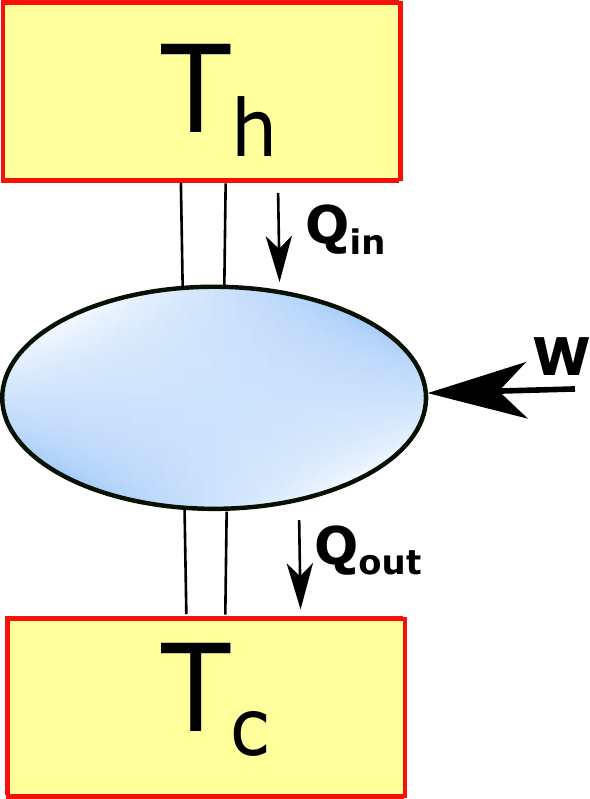}\\ \hline
 \textbf{Heat Engine}& \textbf{Refrigerator} & \textbf{Joule Pump} & \textbf{Cold Pump} \\ \hline
 $Q_{in}$>0, $Q_{out}$<0, & $Q_{in}$<0, $Q_{out}$>0 & $Q_{in}$<0, $Q_{out}$<0 & $Q_{in}$>0, $Q_{out}$<0\\
  $W>0$, $\eta = \frac{W}{Q_{in}}$ &  $W<0$, $COP = \frac{Q_{out}}{|W|}$& $W<0$, $COP = \frac{|Q_{out}|+|Q_{in}|}{|W|}$ & $W<0$, $COP = \frac{|Q_{out}|}{|W|}$ \\ \hline
\end{tabular}}

\caption{The above chart defines various types of regimes possible for a quantum thermodynamic cycle on the basis of the signs of work done and heat exchanged. }
\label{table}
\end{table}
The categorization is summarized here in Fig.~\ref{table}.
In our calculations, we assume that if work is positive, then work done by the system, and if it is negative, it is done by the system. If heat exchanged is positive, then heat is absorbed by the cycle; if it is negative, heat is released.
This categorization helps us in realizing the significance of negative work. We note that the system always operates as a heat engine if the work done is positive.

\subsection{Quantum Carnot Cycle}
Like the classical Carnot cycle, the quantum Carnot cycle is reversible, involving quantum isothermal and adiabatic processes.
For a quantum Carnot cycle to exist, the energy eigenvalues must satisfy a reversibility condition for all states, see Ref.~\cite{quan2005quantum}. The following is a quantum Carnot cycle(QCC) :
\begin{figure}[H]
\begin{center}
\includegraphics[width=6cm, height = 5cm]{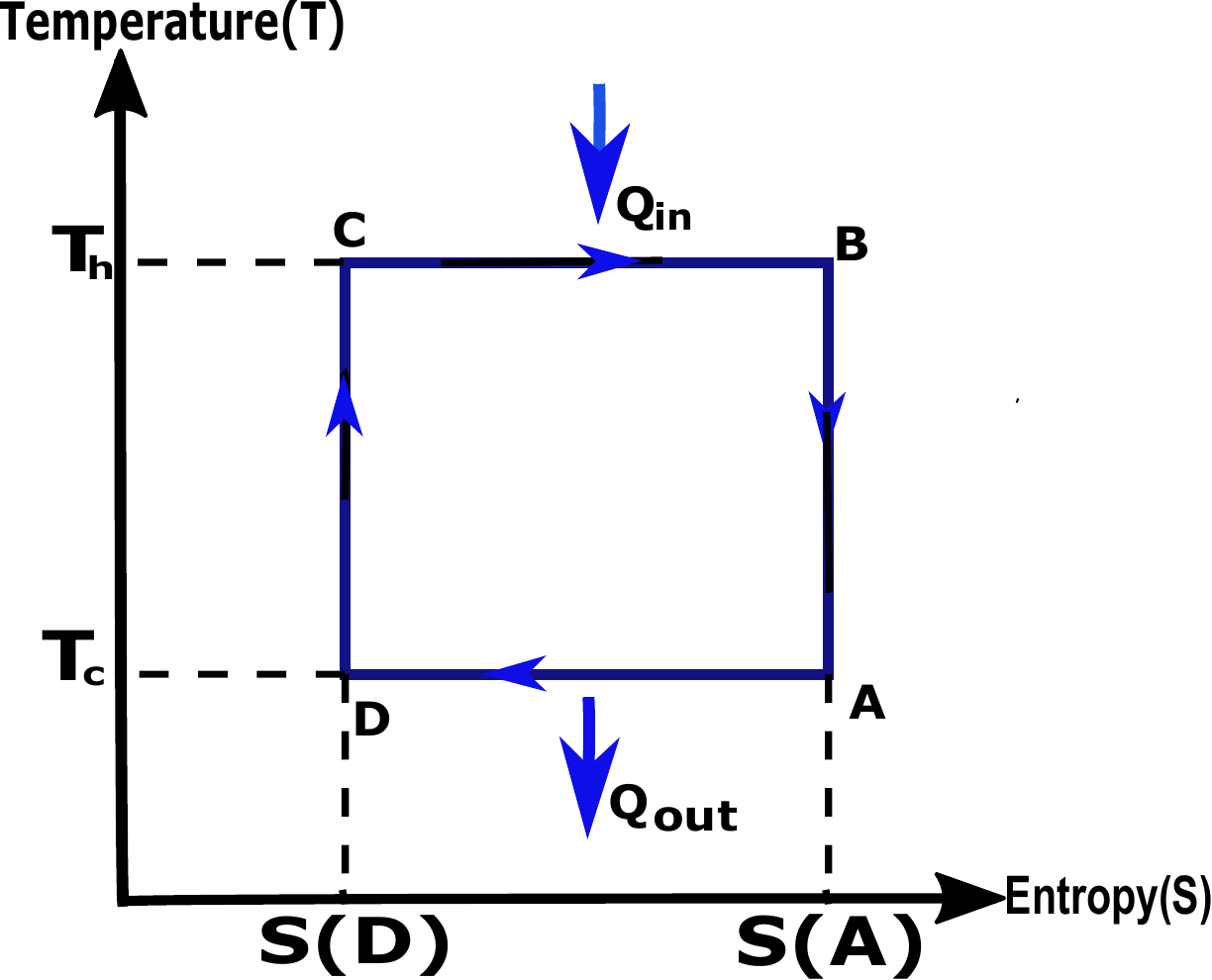}
\end{center}
\caption{Temperature-Entropy curve for QCC. B $\rightarrow$ A and D $\rightarrow$ C are adiabatic processes, C $\rightarrow$ B and A $\rightarrow$ D are isothermal processes.}
\label{fig:numvspert1}
\end{figure}
B $\to$ A and D $\to$ C are adiabatic strokes such that,
\begin{equation}
P_{n}(A) = P_{n}(B) , P_{n}(C) = P_{n}(D),
\end{equation}
which implies,
\begin{equation}
\begin{array}{l}
S(A) = S(B) \mbox{ and } S(C) = S(D).
\end{array}
\end{equation}
Throughout the paper, we use the following notations for convenience:
\begin{enumerate}
\item $p_h$, $p_c$ $\rightarrow$ position of the impurity during strokes $C \to B$ and $A \to D$, in QOC.
\item $T_h$, $T_c$ $\rightarrow$ Temperature of the hot and cold reservoir.
\item $f_h$, $f_c$ $\rightarrow$ strength of the impurity during strokes $ C \to B$ and $A \to D$, in QOC.
\item $L_h$, $L_c$ $\rightarrow$ length of the ISW at C and D.
\item $k_{B}$ $\rightarrow$ Boltzmann constant.
\end{enumerate}
For all our calculations, the temperature of the cold reservoir is fixed at $T_c$ = $1.5K$ and work output as expressed in millielectronvolt (meV) or micro electronvolt ($\mu eV$).
Expressions for heat exchanged are:
\begin{equation}
Q_{\mathrm{in}}=T_{h}[S(B)-S(C)]>0,
\mbox{ and }
Q_{\mathrm{out}}=T_{c}[S(D)-S(A)]<0
\end{equation}
Hence work output is:
\begin{equation}
W_{\mathrm{Carnot}}=Q_{\mathrm{in}}+Q_{\mathrm{out}}=\left(T_{h}-T_{c}\right)[S(B)-S(D)],
\label{WC}
\end{equation}
where $S(i)$ is entropy here in $i \in \{A,B,C,D\}$, and \\
\begin{equation}
S(i) = -k_{b}\sum\frac{exp[-\beta E_{n}(i)]}{Z(i)}[-\beta_{i}E_{n}(i)-\ln Z(i)], \label{entropy}
\end{equation}
and efficiency is
\begin{equation}
\eta_{\mathrm{Carnot}}=\frac{W_{\mathrm{Carnot}}}{Q_{\mathrm{in}}}=1-\frac{T_{c}}{T_{h}},
\end{equation}
which matches with the classical expression for efficiency of QCC. The reversibility condition as derived in Ref. \cite{quan2007quantum}, is given as:
\begin{equation}
E_n(A)-E_m(A)= \frac{T_c}{T_h}[E_n(B) - E_m(B)].
\end{equation}
Substituting the strong coupling energy eigenvalue (Eq.~(\ref{strongeigen})) in the above equation, we get,
\begin{eqnarray}
\hspace{-2.5cm}
&& \frac{\hbar^{2}}{2mL^{2}_{A}}\left[\left(\frac{n^2\pi^2}{p^{2}_{A}} + \frac{nf\pi}{p^{3}_{A}}\right)-\left(\frac{m^2\pi^2}{p^{2}_{A}} + \frac{mf\pi}{p^{3}_{A}}\right)\right] \nonumber\\&=&\frac{T_{c}}{T_{h}} \frac{\hbar^{2}}{2mL^{2}_{B}}\left[\left(\frac{n^2\pi^2}{p^{2}_{B}} + \frac{nf\pi}{p^{3}_{B}}\right)-\left(\frac{m^2\pi^2}{p^{2}_{B}} + \frac{mf\pi}{p^{3}_{B}}\right)\right],
\end{eqnarray}
where $L_{i}$ and $p_{i}$ are length of ISW and position of impurity at $i^{th}$ instant. The only case for which we will get the reversibility condition satisfied independent of the states $m,n$ is when we vary length of the ISW during the cycle and keeping other parameters like position and strength constant during the cycle. Thus, we have $p_{A} = p_{B}$. This gives us,
\begin{equation}
\frac{L_{B}^2}{L_{A}^2} =\frac{T_{c}}{T_{h}}.
\label{34}
\end{equation} The reversibility condition, thus can be satisfied by taking appropriate values of length of ISW as shown in Eq.~(\ref{34}). The cycle will be analysed for numerous cases by varying different parameters for the strength and position of the impurity. Let us calculate the work output and efficiency of QCC when we consider a particle in infinite square well without impurity. In that case the energy eigenvalues have form $E_{n}(i) = \gamma_{i}n^2$ where $\gamma_{i} = \frac{h^2}{8mL^2}$. The entropy of the working substance can be calculated by using Eq.~(\ref{entropy}), we get:
$S(i) = \frac{k_{b}}{2} + k_{b} \ln(\frac{1}{2}\sqrt{\frac{\pi}{\beta_{i}\gamma_{i}}})$ at $i^{th}$ instant.
While deriving this, we used the approximation $\sum_{n=1}^{\infty} \exp (-\beta E_{n}) \approx \int_{0}^{\infty} \exp (-\beta E_{n}) dn$. Let $S_{h}$ and $S_{c}$ be the entropies of the system during strokes BA and DC respectively, which implies $S_{h} = S(A) = S(B)$ and $S_{c} = S(C) = S(D)$.
So the work done during QCC for infinite square well without impurity is,
\begin{equation}
W_{Carnot} = (T_{h}-T_{c})(S_{h} - S_{c}) = k_{b}(T_{h}-T_{c})\ln(\sqrt{\frac{\beta_{c}\gamma_{c}}{\beta_{h}\gamma_{h}}}).
\end{equation}
\subsubsection*{Work output and efficiency of quantum Carnot Cycle in strong coupling regime of impurity}
Using the energy eigenvalue in Eq.~(\ref{strongeigen}) and entropy in Eq.~(\ref{entropy}), we calculate entropies for Carnot cycle up to first order of impurity strength $(f)$ in strong coupling limit. We consider the case when length of ISW is varied during the cycle between $L_{h}$ (length of ISW at instant C) and $L_{c}$ (length of ISW at instant D). Denoting $ S(A)=S(B)=S_{h}$ and $ S(C)=S(D)=S_{c}$, we have,
\begin{eqnarray}
\hspace{-0.9cm}
S_h = \frac{2\beta_{h} k_{b}\sqrt{
\gamma_{h}\beta_{h}} }{\sqrt{p^2\pi}}\left(\frac{1}{2\beta_h} + 2\beta_{h}^{1/2}{\frac{\gamma_{h}^{3/2}}{\pi^{3/2}}}\frac{fg(\gamma_{h})}{p^4}\right)\nonumber + k_{b}\ln{ \left(\frac{1}{2}\sqrt{\frac{{p^2\pi}}{{\beta_h \gamma_{h}}}}\right)},
\end{eqnarray}
and,
\begin{eqnarray}
\hspace{-0.9cm}
S_c = \frac{2\beta_{c} k_{b}\sqrt{
\gamma_{c}\beta_{c}} }{\sqrt{p^2\pi}}\left(\frac{1}{2\beta_c} + 2\beta_{c}^{1/2}{\frac{\gamma_{c}^{3/2}}{\pi^{3/2}}}\frac{fg(\gamma_{c})}{p^4}\right) \nonumber+ k_{b}\ln{\left(\frac{1}{2}\sqrt{\frac{{p^2\pi}}{{\beta_c \gamma_{c}}}}\right)}.
\end{eqnarray}
The work done during QCC is,
\begin{equation}
W_{Carnot} = (T_{h}-T_{c})(S_{h} - S_{c}).
\end{equation}
Substituting expressions for $S_{h}$ and $S_{c}$ in the above equation for work done we get,
\begin{eqnarray*}
\hspace{-1.4cm}
W_{Carnot} &= & (T_{h}-T_{c})\frac{2k_{b}}{\sqrt{p^2\pi}}\beta_{h}^{3/2}
\gamma_{h}^{1/2}\bigg(\frac{1}{2\beta_h} + 2\beta_{h}^{1/2}{\frac{\gamma_{h}^{3/2}}{\pi^{3/2}}}\frac{fg(\gamma_{h})}{p^4}\bigg)\nonumber\\ &-& \beta_{c}^{3/2}
\gamma_{c}^{1/2} \bigg(\frac{1}{2\beta_c}+ 2\beta_{c}^{1/2}{\frac{\gamma_{c}^{3/2}}{\pi^{3/2}}}\frac{fg(\gamma_{c})}{p^4}\bigg)+ \frac{\sqrt{p^2\pi}}{2}\ln{\bigg(\sqrt{\frac{\beta_{c}\gamma_{c}}{\beta_{h}\gamma_{h}}}}\bigg)
\end{eqnarray*}
and for efficiency,
\begin{equation}
\eta_{Carnot}=\frac{W_{Carnot}}{Q_{in}}=\frac{(T_{h}-T_{c})(S_{h}-S_{c})}{T_{h}(S_{h}-S_{c})}=1-\frac{T_c}{T_h},
\end{equation}
wherein
$g(\gamma_{i}) = \frac{2p^2-f\sqrt{\frac{\beta_{i}\gamma_{i}}{\pi}}}{4\beta_{i}\gamma_{i}}, \gamma_{i} = \frac{h^{2}}{8mL_{i}^{2}}, \mbox {and } \beta_{i} = \frac{1}{k_{b}T_{i}}$ with $i\in \{c,h\}$.
The coefficient of performance (COP) of quantum Carnot refrigerator (QCR) and quantum Carnot cold pump then is,
\begin{equation}
COP_{Carnot}=\frac{Q_{out}}{|W_{Carnot}|}=\frac{T_{c}(S_{h}-S_{c})}{|T_{h}-T_{c}|(S_{h}-S_{c})} =\frac{T_{c}}{|T_{h}-T_{c}|}.
\end{equation}
\subsection{Quantum Otto Cycle}
$4$-strokes of a quantum Otto cycle involve two quantum isochoric and two quantum adiabatic strokes. The quantum analogue of a classical Otto cycle for two-level and multilevel systems is very well discussed in a series of papers, see Refs.~\cite{quan2007quantum,quan2005quantum}. These papers arrive at the same $T-S$ diagram for both classical and quantum Otto cycles. The following is a quantum Otto cycle(QOC),
\begin{figure}[H]
\centering
\includegraphics[width = 6cm, height = 5cm]{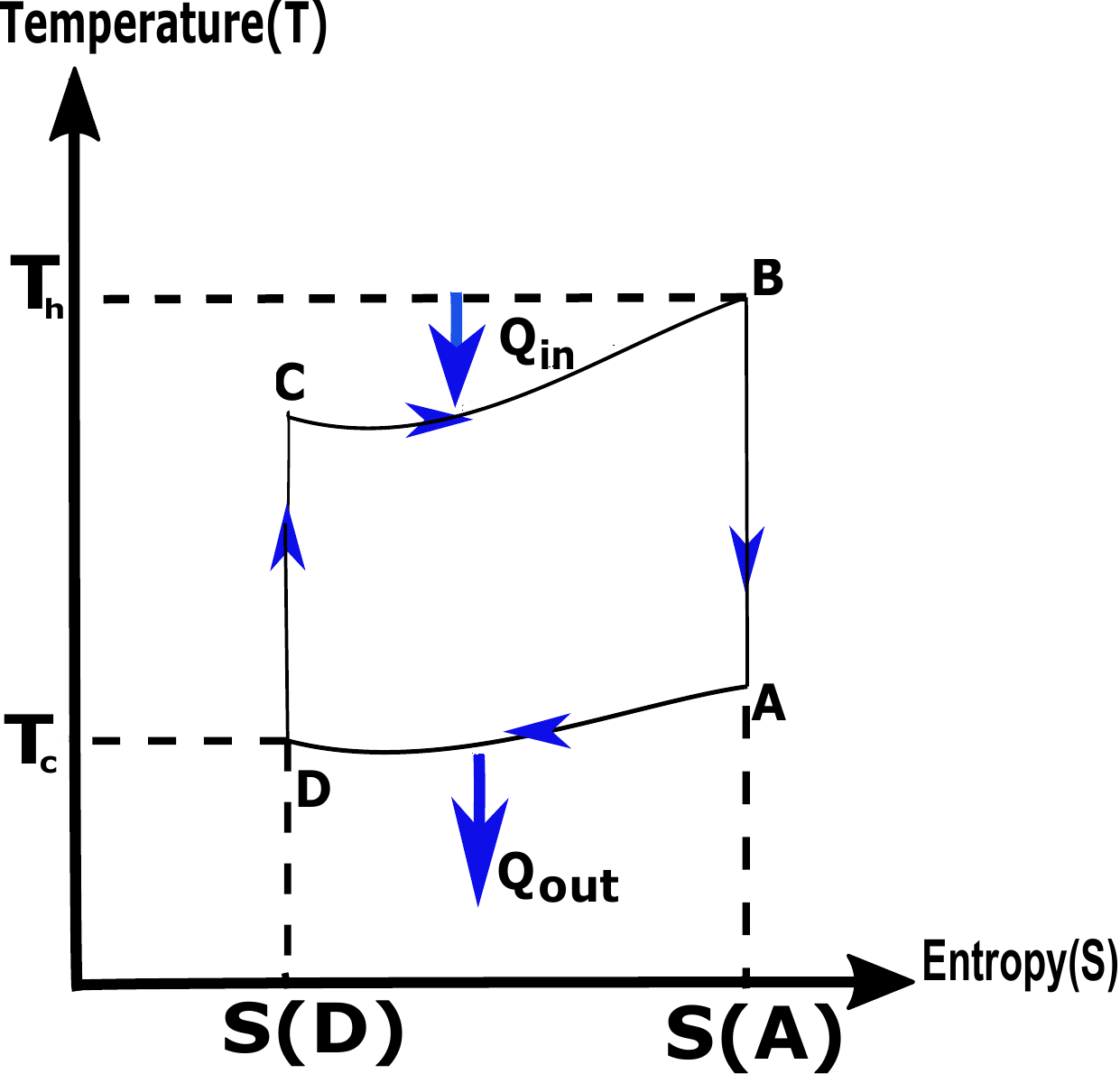}
\caption{Temperature-Entropy curve for QOC. $S(B)$ and $S(D)$ are entropies at temperatures $T_{h}$ and $T_{c}$ respectively. B $\rightarrow $ A and D $\rightarrow$ C are adiabatic processes, C $\rightarrow$ B and A $\rightarrow$ D are isochoric processes.}
\label{fig:my_label}
\end{figure}
Reversibility condition is not required for QOC.
QOC will be evaluated for our impurity model by changing various parameters.
In quantum isochoric processes, the energy eigenvalue of the system remains constant. Since C $\to$ B and A $\to$ D are quantum isochoric strokes:
\begin{equation}
E(B)=E(C)=E^{in}\mbox{ and }E(D)=E(A)=E^{out}.
\end{equation}
B $\to$ A and D $\to$ C are quantum adiabatic strokes during which the occupation probabilities of energy eigenstates remain unchanged, as a result we have,
\begin{equation}
P_{n}(A)=P_{n}(B)\mbox{ and }P_{n}(C)=P_{n}(D).
\end{equation}
The input and output heat can be calculated using Eq.~(\ref{prime}) by \hspace{1cm}taking an integral over path $B \to A$ and path $D \to C$ respectively, as was done in Ref.~\cite{quan2007quantum}. In C $\to$ B stroke, the entropy of the system increases, more heat is absorbed by the system hence $Q_{in}>0$, similarly entropy of the system decreases during the stroke A$\to$D which results in $Q_{out}<0$ that is heat is released by the system.
Following are the expressions for heat exchanged for QOC,
\begin{eqnarray}
Q_{in}&=&\int_{C}^{B} E_{n} \,dP_{n} = \sum_{n}E_{n}^{in}(P_{n}(B) - P_{n}(D))>0,\label{Qin}\\
Q_{out}&=&\int_{A}^{D} E_{n} \,dP_{n} =\sum_{n}E_{n}^{out}(P_{n}(D) - P_{n}(B))<0.\label{Qout}
\end{eqnarray}
The work performed by QOHE~\cite{quan2007quantum} is then,
\begin{equation}
W_{Otto}=Q_{in} + Q_{out}=\sum_{n}(E_{n}^{in}-E_{n}^{out})(P_{n}(B)-P_{n}(D)),
\label{WO}
\end{equation}
and the efficiency for QOHE is,
\begin{equation}
\eta_{O}=\frac{Q_{in}}{W_{Otto}} = \frac{\sum_{n}E_{n}^{in}(P_{n}(B) - P_{n}(D)}{\sum_{n}(E_{n}^{in}-E_{n}^{out})(P_{n}(B)-P_{n}(D))}.
\end{equation}

Let us analyse what happens when we have a ISW potential with no impurity such that the length of the ISW varies from $L_{c}$(length of square well during stroke $A \to D$) to $L_{h}$(length of square well during stroke $C \to B$). The eigenvalues and eigenfunctions for ISW potential are of form $E_{n}(i) = \gamma_{i}n^2$ where $\gamma_{i} = \frac{(h)^2}{8mL_{i}^2}$. We use the approximation as stated in ~\cite{PhysRevE.86.061108,quan2007quantum} with $\sum_{n=1}^{\infty} \exp (-\beta E_{n}) \approx \int_{0}^{\infty} \exp (-\beta E_{n}) dn$ and this gives an expression for work output and efficiency for system without impurity, that is ISW without Dirac delta potential.
\begin{equation}
W_{Otto}\approx\frac{1}{2}(\gamma_{h}-\gamma_{c})\bigg(\frac{1}{\beta_{h}\gamma_{h}}-\frac{1}{\beta_{c}\gamma_{c}} \bigg), \quad \textrm{and} \quad
\eta_{Otto} \approx 1-\frac{\gamma_{c}}{\gamma_{h}},\label{noimpeffo}
\end{equation}
where $\gamma_{i} = \frac{h^{2}}{8mL_{i}^{2}}$ and $i$ $\in$ $\{c, h\}$.
\subsubsection*{Work output and efficiency of quantum Otto Cycle in strong coupling regime of impurity}
Using the energy eigenvalue in Eq.~(\ref{strongeigen}) and heat equations in Eqs.~(\ref{Qin}),(\ref{Qout}) we calculate $Q_{in}$ and $Q_{out}$ for Otto cycle up to first order in strength parameter$(f)$ in the strong coupling limit. We consider the case when the position of the impurity is varied during the cycle between as $p_{h}$ (position of impurity during stroke $A \to D$) and $p_{c}$ (position of impurity during stroke $C \to B$),
\begin{eqnarray}
\hspace{-0.8cm}
Q_{in} = \frac{2f\gamma^{3/2}}{p_h^{3}\pi^{3/2}}\bigg(\frac{g(p_h)\beta_h^{1/2}}{p_h}-\frac{g(p_c)\beta_c^{1/2}}{p_c}\bigg) +\frac{1}{2}\bigg(\frac{1}{\beta_{h}}-\frac{p_c^2}{\beta_{c}p_h^2}\bigg),
\end{eqnarray}
\begin{eqnarray}
\hspace{-0.8cm}
Q_{out} = \frac{2f\gamma^{3/2}}{p_c^{3}\pi^{3/2}}\bigg(\frac{g(p_c)\beta_c^{1/2}}{p_c}-\frac{g(p_h)\beta_h^{1/2}}{p_h}\bigg) +\frac{1}{2}\bigg(\frac{1}{\beta_{c}}-\frac{p_h^2}{\beta_{h}p_c^2}\bigg),
\end{eqnarray}
where
$ g(p_{i}) = \frac{2p_{i}^2-f\sqrt{\frac{\beta_{i}\gamma}{\pi}}}{4\beta_{i}\gamma}
$, $\gamma = \frac{h^{2}}{8mL^{2}}$ and $\beta_{i} = \frac{1}{k_{b}T_{i}}$, such that $i\in \{c,h\}$.
Using the expressions for $Q_{in}$, $Q_{out}$ we get the work done up to first order in strength parameter($f$) as,
\begin{eqnarray}
\hspace{-0.8cm}
&&W_{Otto} = \frac{2f\gamma^{3/2}}{\pi^{3/2}}\bigg(\frac{g(p_h)\beta_h^{1/2}}{p_h}-\frac{g(p_c)\beta_c^{1/2}}{p_c}\bigg)\bigg(\frac{1}{p_h^3}-\frac{1}{p_c^3}\bigg)\nonumber\\&+&\bigg(\frac{p_h^2-p_c^2}{2}\bigg)\bigg(\frac{1}{\beta_{c}p_h^2}-\frac{1}{\beta_{h}p_c^2}\bigg).
\end{eqnarray}

To express efficiency $\eta =\frac{W_{Otto}}{Q_{in}}$ in a simplified manner, let $x =\frac{2f\gamma^{3/2}}{\pi^{3/2}}\Big(\frac{g(p_h)\beta_h^{1/2}}{p_h}-\frac{g(p_c)\beta_c^{1/2}}{p_c}\Big)$ and $y= \frac{\beta_{h}p_{c}^2- \beta_{c}p_{h}^2}{2\beta_{h}\beta_{c}}$, thus
\begin{equation}
\eta_{Otto} = 1 - \bigg(\frac{p_{h}^3}{p_{c}^3}\bigg)\bigg(\frac{x-yp_{c}}{yp_{h}-x}\bigg).
\end{equation}
Similarly, the coefficient of performance, $ COP =\frac{|Q_{out}|}{|W|}$ for quantum Otto cold pump (QOCP) and quantum Otto refrigerator (QOR) is,
\begin{equation}
COP_{Otto} = \frac{p_{h}^3(yp_{c}-2x)}{2x(p_{c}^3-p_{h}^3)+yp_{h}p_{c}(p_{h}^2-p_{c}^2).}
\end{equation}
Unlike in strong coupling, we could not get closed-form expressions of work output in the weak coupling limit due to the non-trivial terms in the weak coupling energy eigenvalue, which make the integrals in work done diverge.

\section{Results}\label{results}

In this section, we plot work done and efficiency for QOC and QCC when changing various parameters. This section is divided into four subsections covering the cases of varying strength, length, and position of the impurity \emph{during} the cycle for strong and weak coupling regimes. As the analytical expression for work and efficiency could be found only for strong coupling, we will numerically analyze the cases of weak coupling. To generate the density plots of work done and efficiency, we used our derived energy eigenvalues in Eqs.~(\ref{strongeigen}),~(\ref{perturb}) and calculated the summations in Eqs.~(\ref{entropy}),~(\ref{Qin}),~(\ref{Qout}). Wolfram Mathematica file for generating plots for the case of varying impurity strength during the cycle has been uploaded to  \href{https://github.com/Aditya214/Wolfram-Mathematica-code-for-varying-strength-of-the-impurity-during-the-cycle-in-weak-coupling-.git}{github}.\footnote{We have openly released our Wolfram Mathematica code in \href{https://github.com/Aditya214/Wolfram-Mathematica-code-for-varying-strength-of-the-impurity-during-the-cycle-in-weak-coupling-.git}{github} for the case of varying the strength of impurity during the cycle. 
}

In the first three subsections, we discuss the results for weak coupling, and then similar cases are briefly discussed for strong coupling in the last subsection.

\subsubsection*{Variation of Parameters}\label{variation}
The variable parameters for our impurity model are the strength of impurity ($f$), position of impurity ($p$), and length of the ISW ($L$). It is important to note that the parameters will be varied in two ways:
\begin{enumerate}
\item Changing a particular parameter \emph{during} the cycle, this means the parameter changes its values while a cycle is going on.
\item Changing parameter cycle wise, meaning the particular parameter is constant \emph{during} the cycle and then changes its value in the next cycle.
\end{enumerate}
Thus, we will have, in total, three possible cases,
\begin{enumerate}
\item Changing strength \emph{during} the cycle, with other parameters varying after every cycle. (QCC does not exist for this case as the reversibility condition is not satisfied)
\item Changing length \emph{during} the cycle, with other parameters varying after every cycle. (QCC exists for this case).
\item Changing position \emph{during} the cycle, with other parameters varying after every cycle. (QCC does not exist for this case as the reversibility condition is not satisfied).
\end{enumerate}
QOC does not require any reversibility condition; it exists for all three cases.

\subsection{Changing strength of the impurity during the cycle for weak coupling.}\label{IIIA}
\begin{figure}
\begin{center}
\includegraphics[width=6cm, height = 5cm]{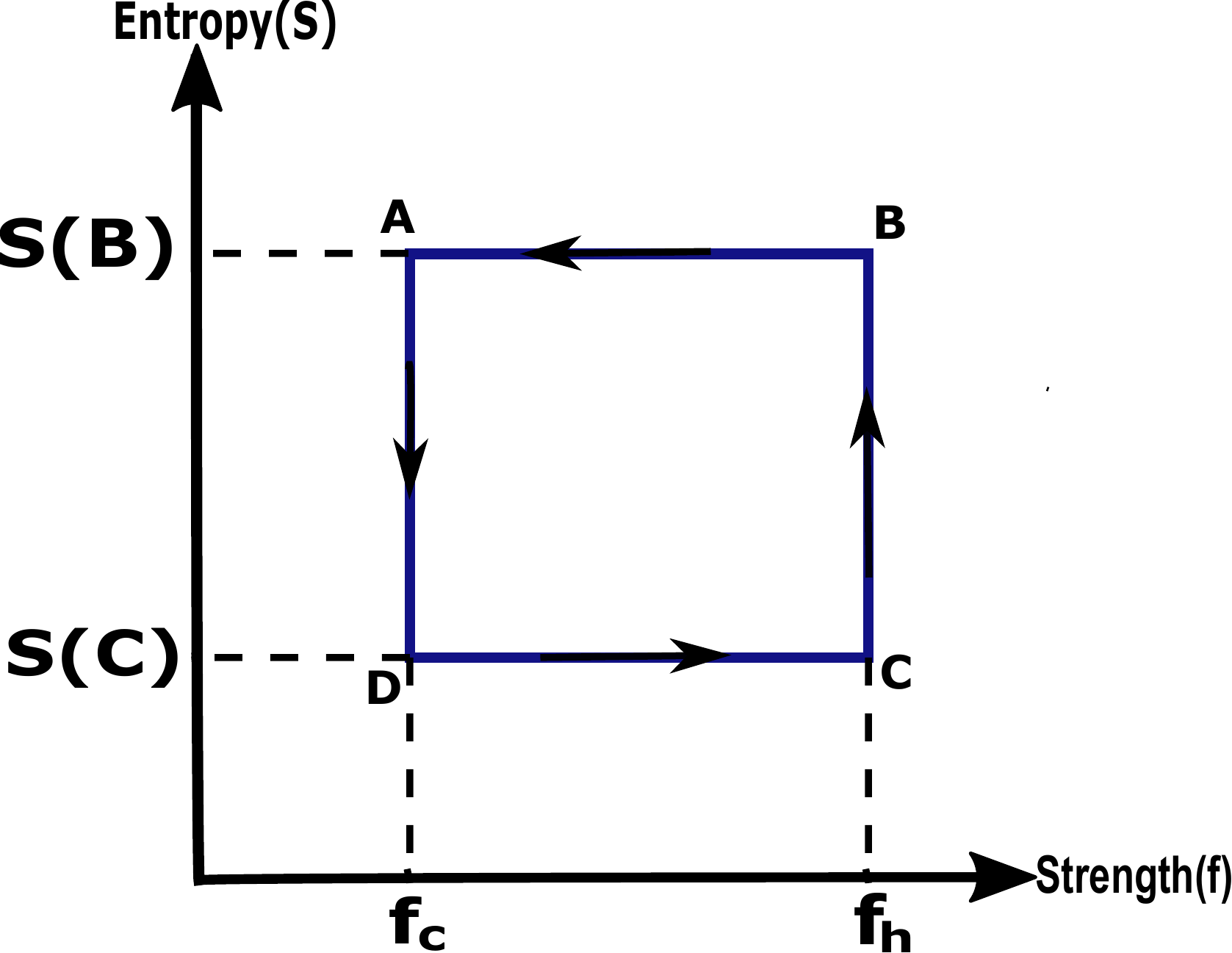}
\end{center}
\caption{ Quantum Otto cycle operating between a hot bath at fixed temperature $T_{h}$ and a cold bath at fixed temperature $T_{c}$. It has two adiabatic (B$\rightarrow$ A and D$\rightarrow$ C) and two isochoric (C$\rightarrow$ B and A$\rightarrow$ D) strokes. Strength changes during the cycle between $f_{h}$ and $f_{c}$}.
\label{fig:Otto}
\end{figure}

In this subsection, the impurity's strength ($f$) will be varied \emph{during} the cycle. In contrast, other parameters such as the temperature of the hot reservoir ($T_h$), length of the well ($L$), and position of the impurity ($p$) will be constant \emph{during} the cycle. Fig.~\ref{fig:Otto} shows one complete cycle of QOC where we vary the strength along strokes $B \to A$ and $D \to C$ for a fixed length $L$ of the well and fixed position $p$ of the impurity.

\begin{figure}
\includegraphics[width = 9cm, height = 9cm]{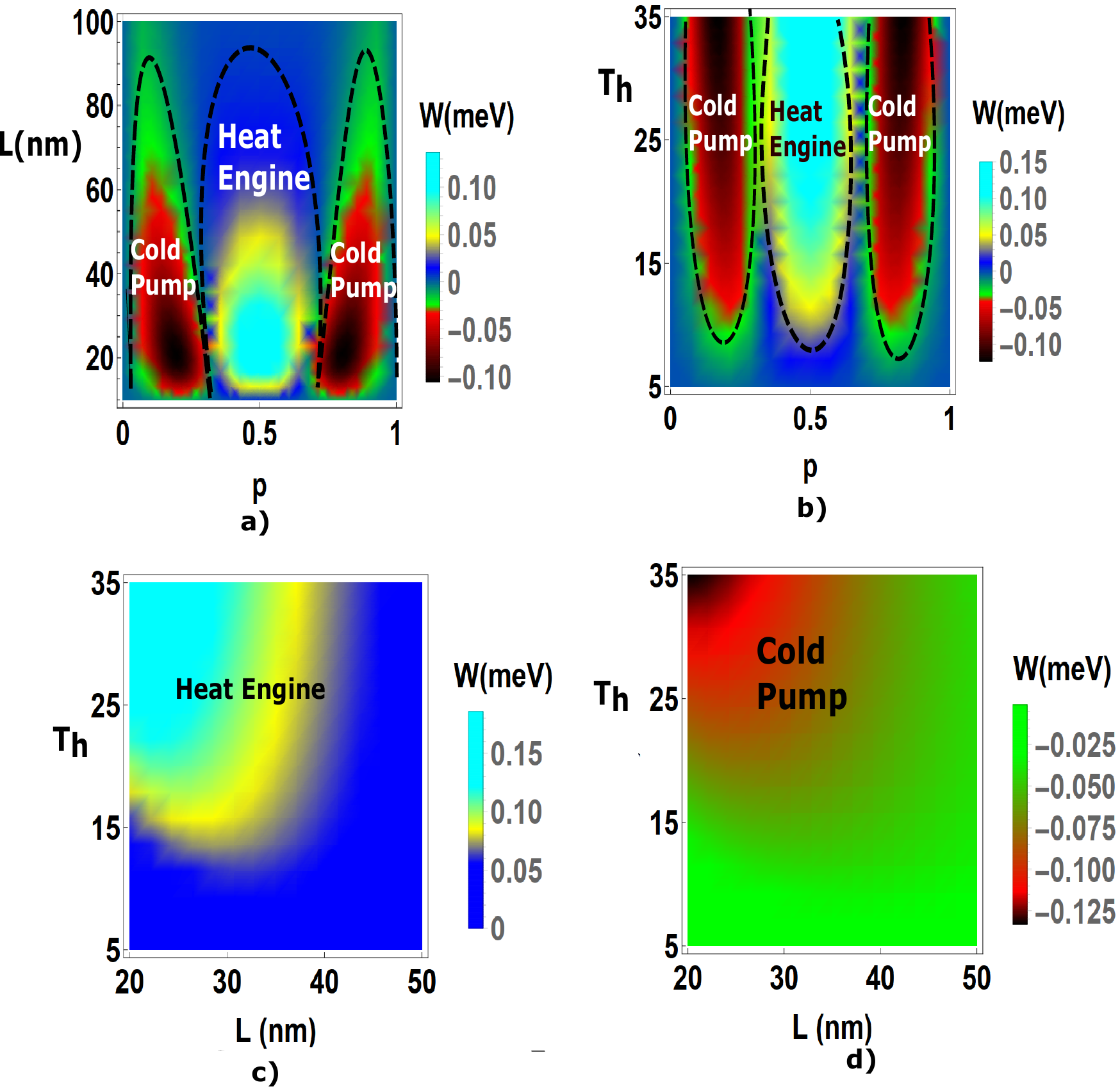}
\caption{Density plots for work output in meV of QOC. \textbf{a)}$T_{h} =25K$, \textbf{b)}$L =25nm$, \textbf{c)} $p=0.5$, \textbf{d)} $p =0.15$.}
\label{fig8}
\end{figure}

The work performed in QOC (Eq.~(\ref{WO})) is,
\begin{equation}
W_{O} = \sum_{n}[E_{n}(f_{h})-E_{n}(f_{c})][P_{n}(B)-P_{n}(D)],
\end{equation}
where $E_{n}(f_{h}), E_{n}(f_{c})$ are the $n^{th}$ energy levels associated with the two isochoric processes since we know that no work is done in an isochoric or isoenergetic process which implies constant energy levels in accordance with Eq.~(\ref{prime}. For adiabatic processes the heat exchanged is zero and hence we get constant occupation probability for each energy level in accordance with Eq.~(\ref{prime}). When strength of the impurity is varied \emph{during} the cycle it is noted that the Carnot reversibility condition is not satisfied, hence QCC does not exists for this case.
Unlike QCC we do not require the reversibility condition for the QOC Ref.~\cite{quan2007quantum}. We expect the work output to be symmetric around, $p = 1/2$ since the eigenvalues are symmetric around $p =1/2$ as $E_{n}(f,p) = E_{n}(f, 1-p)$. The work output should approach the work output of ISW without impurity if the impurity is very close to the wall. An impurity close to the wall implies zero work. Entropy $S$ varies between $S(T_{c})$ and $S(T_{h})$. We tuned the strength $f_{h}$, $f_{c}$ and we found different operational phases of the system existing for $f_{h} = 1, f_{c} = -1$. In case of ISW without impurity if we keep the length of the well fixed during one complete Otto cycle then $W_{O} = 0$ which can be easily seen from Eq.~(\ref{noimpeffo}) and this implies $\eta_{O} = 0$.

\begin{figure}[H]
\includegraphics[width = 9cm, height =9cm]{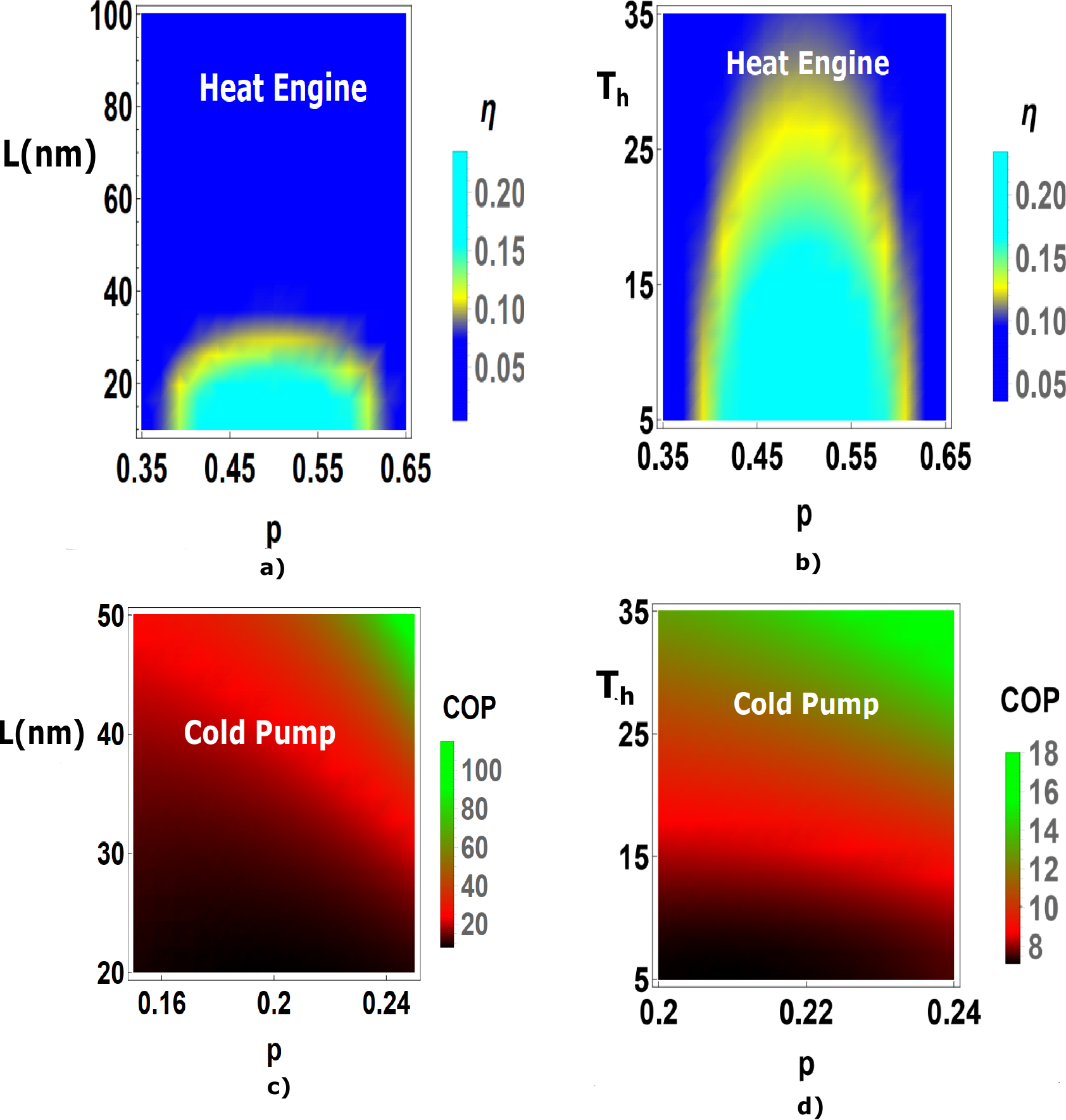}
\caption{Efficiency and coefficient of performance of QOC. \textbf{a)} $T_{h}$ =25K, \textbf{b)} L =25nm, \textbf{c)} $T_{h}$ = 25K, \textbf{d)} L=25nm. }
\label{figure9}
\end{figure}

Fig.~\ref{fig8}(a) shows two operational phases, QOHE and QOCP. The system works as heat engine for the impurity position $0.35<p< 0.65$ and works as cold pump for $0.16<p<0.3$ and $0.7<p<0.9$. Similarly, Fig.~\ref{fig8}(b) shows two operational phases, QOHE and QOCP. In Fig.~\ref{fig8}(c) there is no negative work done, hence the QOC operates as QOHE. In the Fig.~\ref{fig8}(d) the system functions as QOCP. Figs.~\ref{figure9}(a),\ref{figure9}(b) are the efficiency plots of the QOHE corresponding to Figs.~\ref{fig8}(a),\ref{fig8}(b). Figs.~\ref{figure9}(c),\ref{figure9}(d) show coefficient of performance (COP) for cold pump corresponding to Figs.~\ref{fig8}(a) and \ref{fig8}(b) respectively. Note efficiency is plotted in Fig.~\ref{figure9}(a) for the impurity position $0.35<p< 0.65$ which is the region where the system behaves as QOHE and similarly the COP of cold pump corresponding to Fig.~\ref{fig8}(b) is shown in Fig.~\ref{figure9}(c) for $0.16<p<0.24$.

\vspace{-0.5cm}
\subsection{Changing length during the cycle for weak coupling}\label{IIIB}
In this regime, Carnot reversibility condition is satisfied while changing length of the ISW during the cycle. The Carnot reversibility condition is given as,~$E_n(D) - E_m(D) = \frac{T_c}{T_h}(E_n(C)-E_m(C))$.
To satisfy the reversibility condition, the equation we get after substituting the energy eigenvalue must be independent of the energy eigenstates $m, n$. So after substituting the energy eigenvalue and simplifying, we get,
$\frac{L_{h}^2}{L_{c}^2} = \frac{T_c}{T_h}$.
Thus, the Carnot reversibility condition can be satisfied for all eigenenergies by choosing the appropriate length and temperature values, which fulfill the reversibility condition.
Hence, in this subsection, we see both QCC and QOC.
Figs.~\ref{fig12} and \ref{fig13} are for QOC when length of the ISW is changed during the cycle.
\begin{figure}
\includegraphics[width = 6cm, height = 6cm]{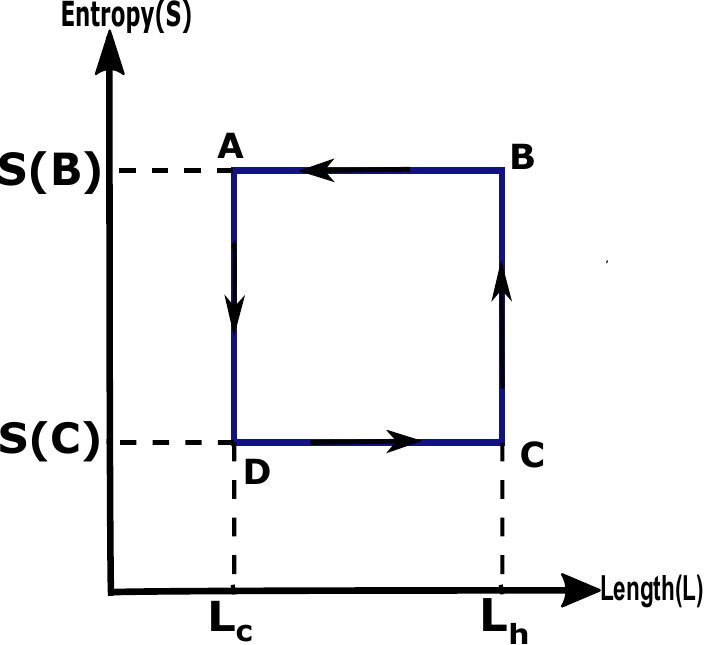}
\centering
\caption{ Quantum Otto cycle for varying length of ISW during the cycle. There are two adiabatic (B $\rightarrow$ A and D $\rightarrow$ C) strokes and two isochoric (C $\rightarrow$ B and A $\rightarrow$ D) strokes.}
\end{figure}
\begin{figure}
\includegraphics[width =8cm , height = 5cm ]{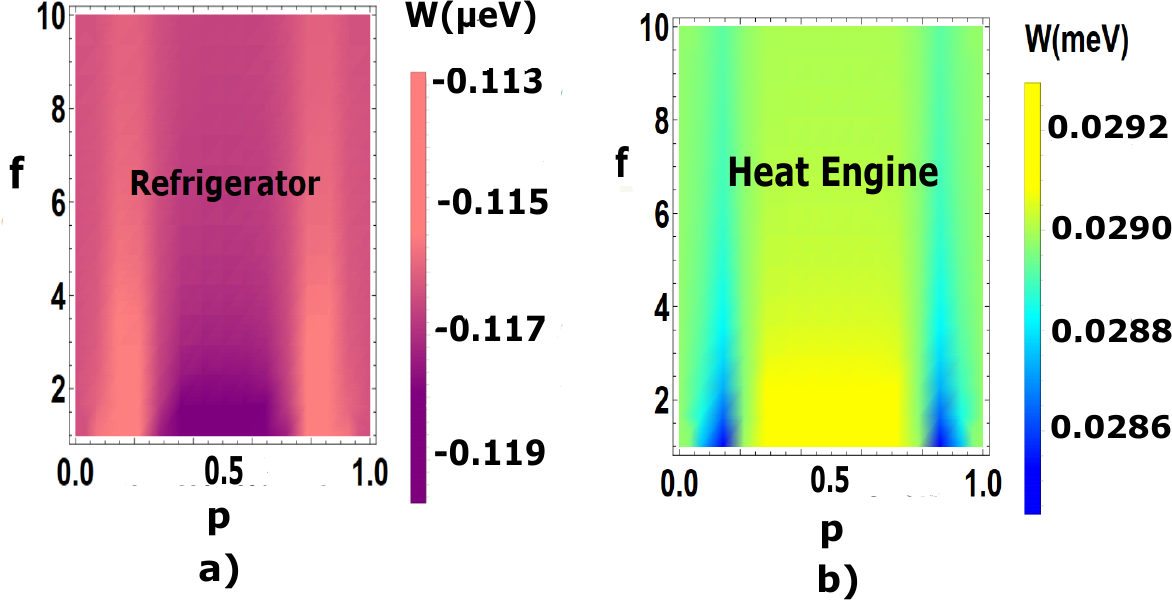}
\caption{Density plots for work output of QOC. \textbf{a)} Parameters: $L_h$ = $100nm$ and $L_c$ = $129nm$, $T_h$ = $2.49K$ (fixed for all cycles), after every cycle position and strength are changed. As the work is entirely negative, the system works as a refrigerator. \textbf{b)} Work is expressed in meV. $L_h$ = $100nm$ and $L_c$ = $163nm$, $T_h$ = $5K$. As the work is positive, the system works as a heat engine.}
\label{fig12}
\end{figure}
A $\rightarrow$ D and C $\rightarrow$ B depict quantum isochoric strokes, hence along them the energy eigenvalue remains constant($dE_n$ = 0). For the energy to remain constant, the length must be kept constant during strokes A $\rightarrow$ D and C $\rightarrow$ B. However, this is not the case with QCC, where length continuously varies throughout the cycle without being constant for any stroke.
\begin{figure}
\includegraphics[width =8cm , height = 5cm ]{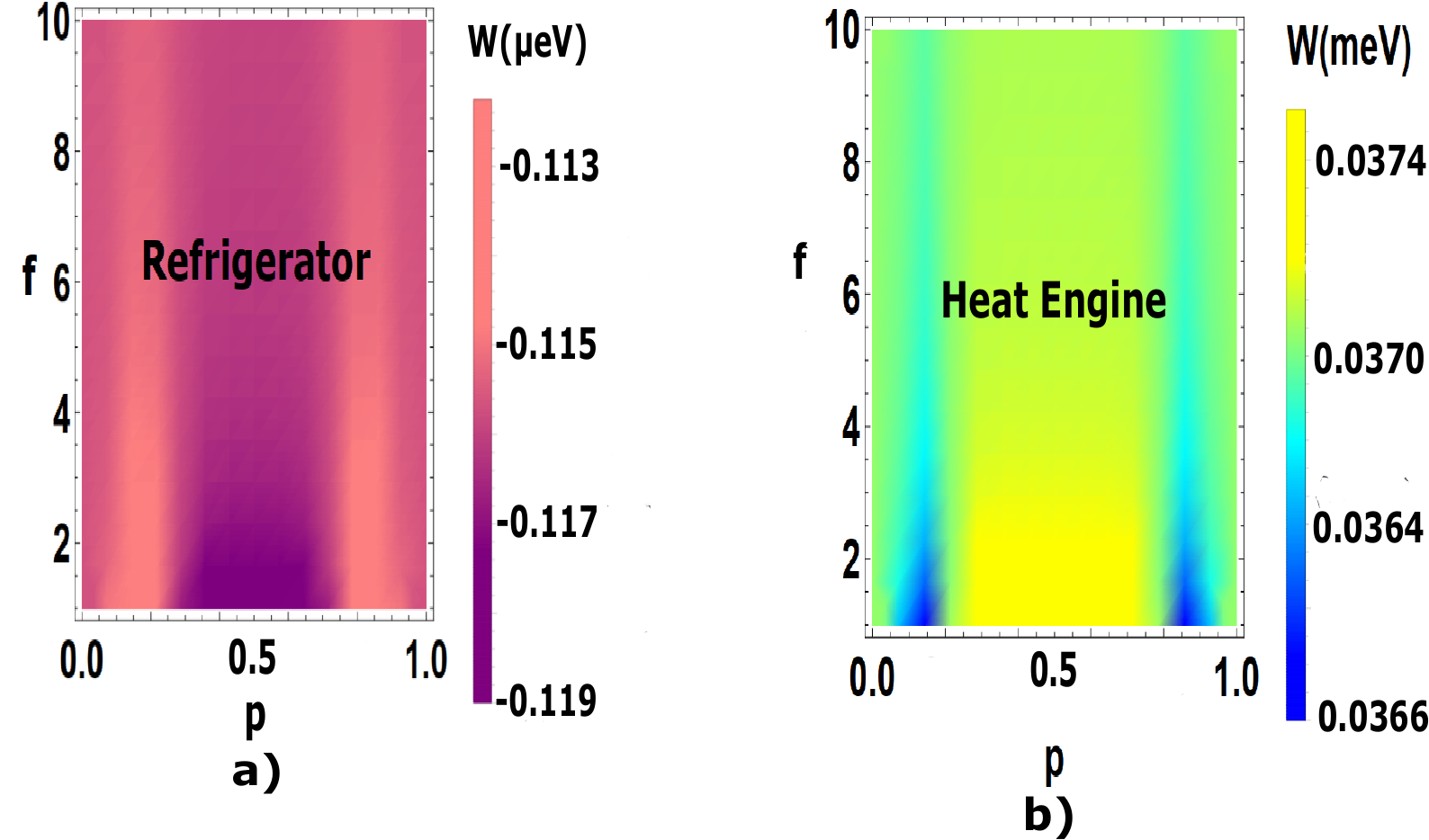}
\caption{Density plots for work output of QCC. \textbf{a)} Parameters: $L_h$ = 100nm and $L_c$ = 129nm, $T_h$ = 2.49K (fixed for all cycles). As work is negative, system works as a refrigerator. \textbf{b)} $L_h$ = 100nm and $L_c$ = 163nm, $T_h$ = 5K. As work is positive, the system works as a heat engine.}
\label{fig13}
\end{figure}
\begin{figure}
{\includegraphics[width=6.5cm,height=5.2cm]{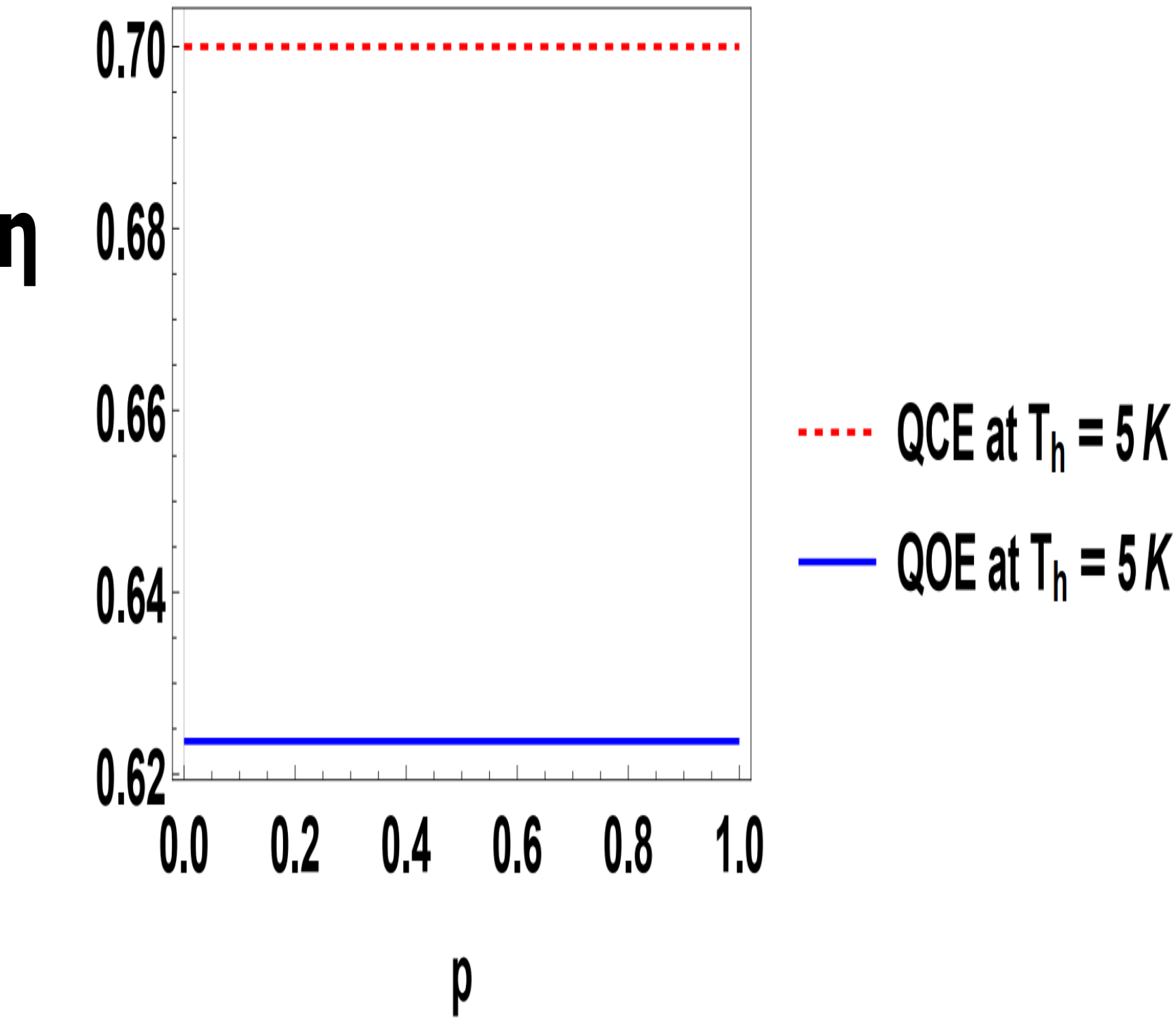} }
\centering
\caption{Plot of efficiency versus position of the impurity in QOHE and QCHE, calculated with $L_h$ = 100nm and $L_c$ = 163nm, $f$ =5 (fixed for all cycles), $T_h$ = 5K(fixed for all cycles).}
\label{2e}
\end{figure}

Now as derived in Eqs.~(\ref{WO}),~(\ref{WC}) we have work done for Otto and Carnot cycles as,
\begin{equation}
W_{Otto} = \sum_{n}[E_{n}(L_{h})-E_{n}(L_{c})][P_{n}(B)-P_{n}(D)].
\end{equation}
$P_n$ being occupation probability,
\begin{equation}
\mbox{and  }W_{Carnot} = (T_{h} -T_{c})[S(B)-S(D)],
\end{equation}
where $S(i)$ is the entropy of the system at $i^{th}$ instant, $i \in$ $\{B,D\}$.
Applying these results, we get work and efficiency plots for both QOC and QCC. Work output plots in Figs.~\ref{fig12}(b),~\ref{fig13}(b) show entirely positive work output, hence the system behaves as a quantum heat engine. Work output is negative in Figs.~\ref{fig12}(a),\ref{fig13}(a). To determine the phase of the system, we analysed the signs of $Q_{in}$ and $Q_{out}$ for those regions. We found that $Q_{in}<0$ and $Q_{out}>0$ from which we conclude that the system operates as a quantum refrigerator. We note that a QCHE produces higher work output and higher efficiency than QOHE.
\subsection{Changing position of the impurity during the cycle for weak coupling}\label{IIIC}
It can be observed that the Carnot reversibility condition is not satisfied for changing position of impurity \emph{during} the cycle. Hence we will analyse only QOC, via varying position during the cycle.
\begin{figure}
\centering
\includegraphics[width=6.5cm, height = 5cm]{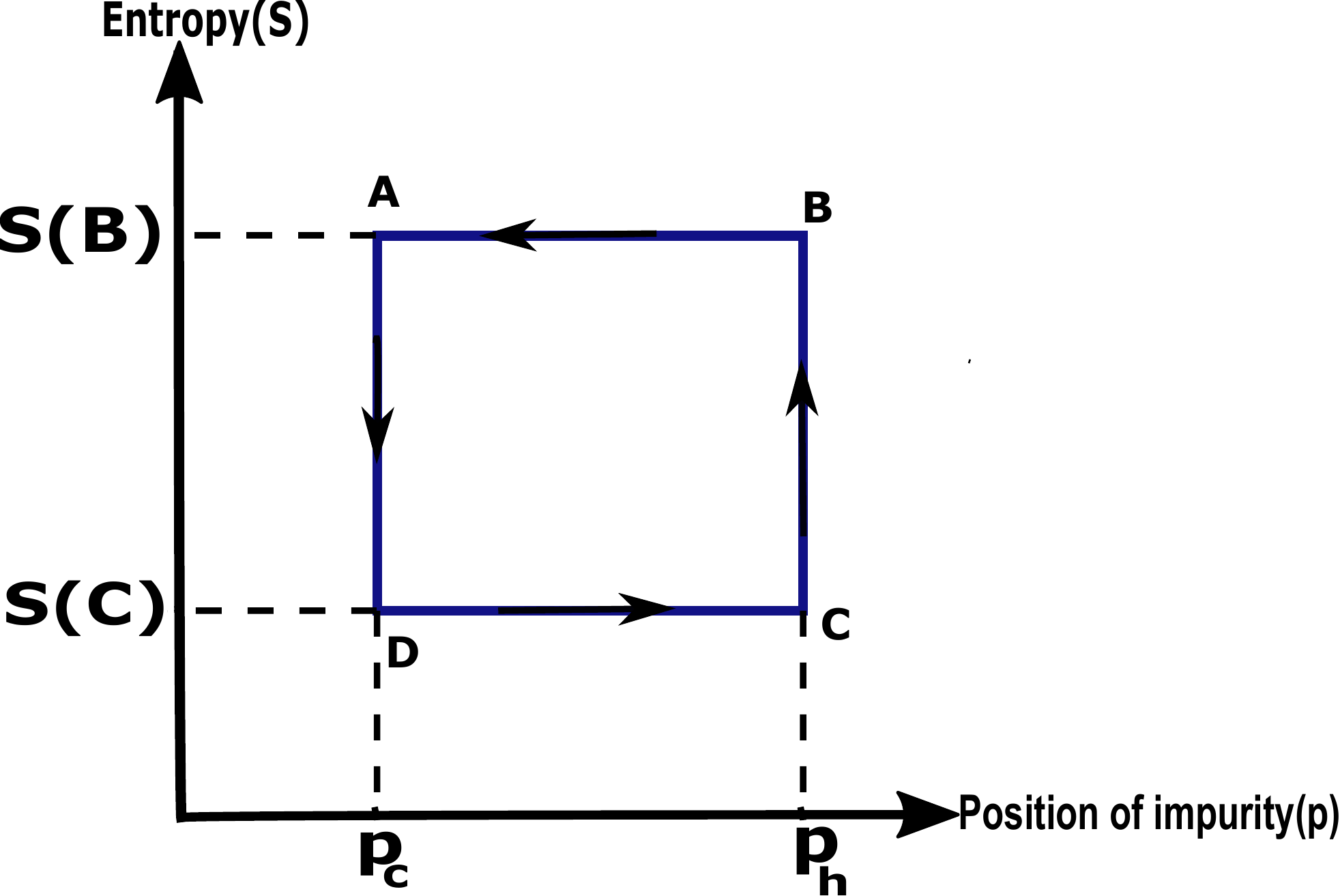}
\caption{Quantum Otto cycle is shown working between $T_{h}$ and $T_{c}$. B $\rightarrow$ A and D $\rightarrow$ C are adiabatic strokes while C $\rightarrow$ B and A $\rightarrow$ D are isochoric strokes. Position of impurity varies between $p_{h}$ and $p_{c}$.}
\label{fig:my_label}
\end{figure}
We calculate the work output of QOHE in this case by using the Eq.~(\ref{WO}), which gives us:
\begin{equation}
W_{Otto} = \sum_{n}[E_{n}(p_{h})-E_{n}(p_{c})][P_{n}(B)-P_{n}(D)].
\end{equation}
\begin{figure}
\includegraphics[width = 8cm,height = 8cm]{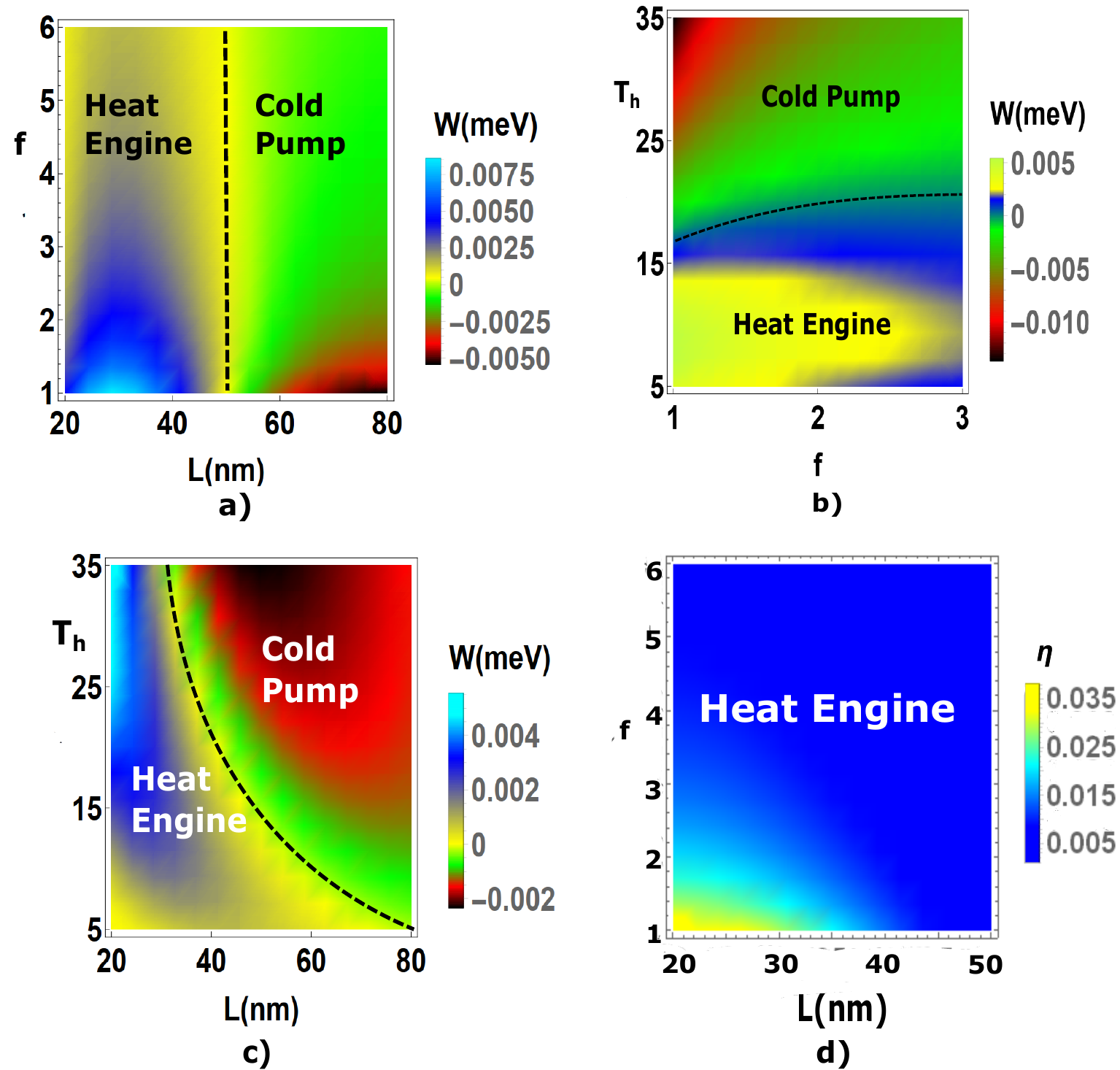}
\caption{Density plots of work for QOC in meV. \textbf{a)} Parameters: $p_h$ = $0.1$ and $p_c$ = $0.8$, $T_h$ = $10K$ (fixed for all cycles), $20nm<L<80nm$ and the strength of impurity varies as $1<f<6$. \textbf{b)} $p_h$ = 0.1, $p_c$ = 0.8 $L$ = 40nm (fixed for all cycles). \textbf{c)} $p_h$ = $0.1$ and $p_c$ = $0.8$, $f$ = $5$ (fixed for all cycles). \textbf{d)} Parameters: $p_h$ = $0.1$ and $p_c$ = $0.8$, $T_h$ = $10K$ (fixed for all cycles), $20nm<L<50nm$ and the strength of impurity varies as $1<f<6$. }
\label{Fig15}
\end{figure}
In Fig.~\ref{Fig15}(a) the system operates as QOHE when the length of ISW is below 50nm, the operational phase of the system changes to QOCP as the length of ISW becomes greater than $50nm$. Similarly, in Figs.~\ref{Fig15}(b), \ref{Fig15}(c) the system operates either as heat engine or as cold pump.
\subsection{Changing position of the impurity during the cycle for strong coupling}\label{IIID}
\begin{figure}
\includegraphics[width = 8cm , height = 8cm]{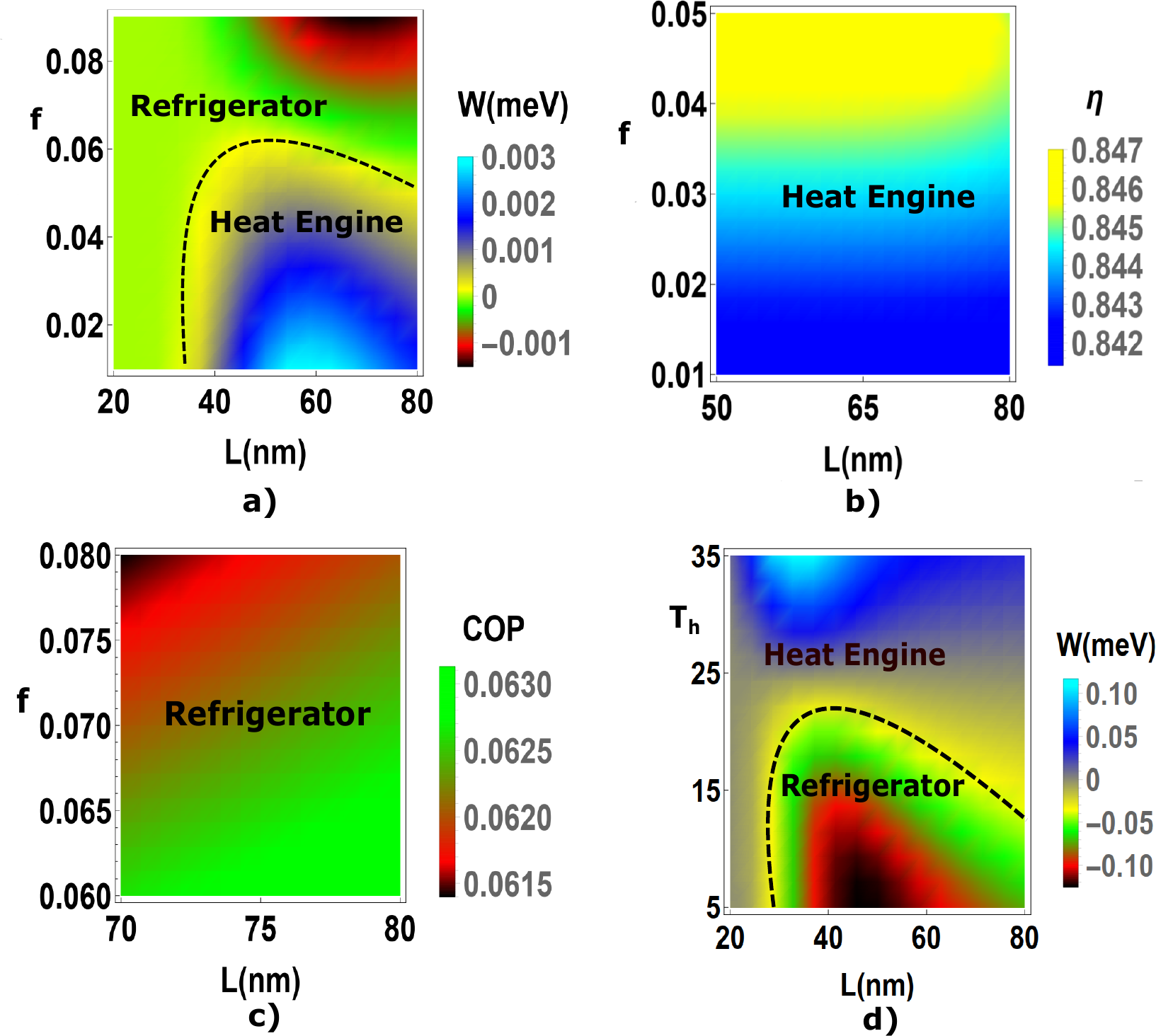}
\caption{Density plots of work and efficiency of QOC. Parameters:\textbf{a)} $p_h$ = $0.2$ and $p_c$ = $0.5$, $T_h$ = $10K$, $20nm<L<80nm$ and the strength of impurity $0.01<f< 0.09$. \textbf{b)} $p_h$ = $0.2$, $p_c$ = $0.5$, $T_h$ = $10K$. \textbf{c)} $p_h$ = 0.2, $p_c$ = 0.5, $T_h$ = 10K. \textbf{d)} $p_h$ = 0.2, $p_c$ = 0.8, $f$ = 0.03.}
\label{Fig17}
\end{figure}

Similar to the weak coupling, we can evaluate our model in strong coupling regime. In this subsection we vary the position of the impurity during the Otto cycle between $p_{h}$ and $p_{c}$ as shown in Fig.~\ref{Fig17}.
In Figs.~\ref{Fig17}(a),~\ref{Fig17}(d), we observe that systems operate either as heat engines or as refrigerators. The previous subsection showed heat engine and cold pump phases when the impurity position varied during QOC for a weak coupling regime. The efficiency of the heat engines obtained for the strong coupling regime is much higher than those obtained in the weak coupling regime, as evident from Figs. \ref{Fig15}(d), \ref{Fig17}(b). Like impurity position, other parameters of our system, such as length of the ISW and strength of the impurity, can also be varied during the cycle for strong coupling, and a similar analysis can be done for them.
\section{Analysis}\label{analysis}

In this section, we analyze the density plots obtained in the \textit{Results} section. We discuss the different thermodynamic operational phases seen in the system and their work output, efficiency, and COP due to embedded impurity in the ISW.

\subsection{Adiabatically varying strength of impurity during Otto cycle in weak coupling limit (\ref{IIIA})}
\vspace{-0.4cm}
While adiabatically varying strength of impurity \emph{during} the QOC cycle, we observe heat engine and cold pump phases.
In Fig.~\ref{fig8}(b), when the length of ISW is fixed at $L =25nm$, we notice that a higher magnitude of work can be obtained by increasing $T_{h}$ for particular values of length.

\begin{table}[H]
\hspace{-1.3cm}
\scalebox{0.73}{
\def\arraystretch{1.5} \small
\begin{tabular}{|l|lll|lll|}

\hline
\multirow{2}{*}
{Adiabatically varying strength ($f_{h}=1$, $f_{c}=-1$)} & \multicolumn{3}{c|}{QOHE   }    & \multicolumn{3}{c|}{QOCP   }    \\ \cline{2-7} 
                           & \multicolumn{1}{l|}{W$_{max}$} & \multicolumn{1}{l|}{$\eta_{max}$} &  W($\eta_{max}$)& \multicolumn{1}{l|}{|W|$_{max}$} & \multicolumn{1}{l|}{$COP_{max}$}
 & |W($COP_{max}$)|\\ \hline
                    $T_h = 25K$, $ 10nm<L<100nm$, $0.35<p<0.65$ & \multicolumn{1}{l|}{0.1}  & \multicolumn{1}{l|}{0.2}& 0.08   & \multicolumn{1}{l|}{0.1}  &\multicolumn{1}{l|}{100}& 0.02     \\ \hline
                    
             $L =25nm$, $5K<T_h<35K$, $0.35<p<0.65$ & \multicolumn{1}{l|}{0.15}  & \multicolumn{1}{l|}{0.2} & 0.10  & \multicolumn{1}{l|}{0.1}  &\multicolumn{1}{l|}{18} & 0.08

          \\ \hline
                 $p= 0.5$, $5K<T_h<35K$, $20nm<L<50nm$ & \multicolumn{1}{l|}{0.15}  & \multicolumn{1}{l|}{0.2} & 0.05 & \multicolumn{1}{l|}{0.125}  &\multicolumn{1}{l|}{35} & 0.05    \\ \hline

\end{tabular}}

\caption{Two different operational phases were revealed while varying strength of the impurity adiabatically. The table shows the magnitude maximum work output in meV, maximum efficiency and maximum COP delivered by the system. Work done here is in $m$eV.}
\label{Table2}
\end{table}

The maximum work output of QOHE and QOCP obtained when the impurity strength is adiabatically varied during the cycle is tabulated in Tables~\ref{Table2}.
$W(COP_{max})$ and $W(\eta_{max})$ are the values of work done at maximum COP and efficiency. The reason these quantities are significant is that the parameter regimes where work output is maximum and where efficiency or COP is maximum may not be the same for some cases. In fact, wherein work output is maximum, efficiency/COP might be low or minimum and wherein work output is less or even minimum, efficiency/COP can be maximum. Thus, the values of $W(COP_{max})$ or $W(\eta_{max})$ could be termed as the effective work output when the thermodynamic cycle is the most efficient.

We use Eq.(\ref{noimpeffo}) to find the work output of the heat engine when we use ISW without impurity. If we compute work using ISW without impurity such that the length of the ISW is kept constant during the cycle, we get zero work output and efficiency. The cold pump phase of the thermodynamic cycle is absent when there is no impurity. Thus introducing an impurity in the ISW has unlocked the cold pump phase in QOC. It also produces non-zero work output in the quantum heat engine phase as if the length is left constant in ISW without impurity.

\subsection{Adiabatically and isothermally varying length of ISW during Carnot/Otto cycle in weak coupling limit (\ref{IIIB})}

\subsubsection{Adiabatically and isothermally varying length of ISW during Carnot cycle}

In contrast to varying strength or position of the impurity during the cycle, the Carnot reversibility condition gets satisfied only while the ISW's length changes. Unlike QOC, the length of ISW in QCC does not remain constant for any of the strokes. Hence, in QCC length of ISW is varied both isothermally and adiabatically.
In Fig.~\ref{fig13}(a), we find that the system absorbs heat from the cold reservoir and releases heat to the hot reservoir, which implies that the system operates as a quantum Carnot refrigerator(QCR). We obtain only heat engine and refrigerator phases while varying the length of ISW during QCC.
\subsubsection{Adiabatically varying length of ISW during Otto cycle}
As the length of ISW remains constant during the isochoric strokes in QOC, length can only be varied adiabatically during the Otto cycle. We obtain negative work output from the signs of $Q_{in}$ and $Q_{out}$. We conclude that the system operates as QOR. Further, in Fig.~\ref{fig12}(b), work done is positive, implying the system acts as a quantum heat engine. Hence, we obtain only heat engine and refrigerator phases while varying the length of ISW during the QOC.
Fig.~\ref{2e} shows that QCHE provides higher efficiency than QOHE.
The maximum values of work outputs, efficiencies of QOC and QCC obtained while varying length of ISW during the cycle are tabulated in Tables \ref{Table3}, \ref{Table4}.

\begin{table}[H]
\hspace{-0.4cm}
\scalebox{0.73}{
\def\arraystretch{1.5} \small
\begin{tabular}{|l|lll|lll|}
\hline
\multirow{2}{*}{Adiabatically varying length ($L_{h}=100nm$, $L_{c}=163nm$)} & \multicolumn{3}{c|}{QOHE  }    & \multicolumn{3}{c|}{QCHE  }    \\ \cline{2-7} 
                           & \multicolumn{1}{l|}{W$_{max}$} & \multicolumn{1}{l|}{$\eta_{max}$}& W(${\eta_{max}}$) & \multicolumn{1}{l|}{W$_{max}$} & \multicolumn{1}{l|}{$\eta_{max}$} & W(${\eta_{max}}$) \\ \hline
                    $T_h = 5K$, $ 1<f<10$, $0<p<1$ & \multicolumn{1}{l|}{29.2}  & \multicolumn{1}{l|}{0.624} &29.2    & \multicolumn{1}{l|}{37.4}  & \multicolumn{1}{l|}{0.7}& 37.4     \\ \hline
                    
             $f=0$ (no impurity) & \multicolumn{1}{l|}{27.2}  & \multicolumn{1}{l|}{0.624}& 27.2    & \multicolumn{1}{l|}{31.8}  &\multicolumn{1}{l|}{0.7} & 31.8

          \\ \hline

\end{tabular}}

\caption{Comparing magnitude of maximum work output and maximum efficiency delivered by the system while acting as quantum heat engine during QOC and QCC. Work done here is in $\mu$eV.}
\label{Table3}
\end{table}


\begin{table}[H]
\hspace{-0.45cm}
\scalebox{0.68}{
\def\arraystretch{1.5} \small
\begin{tabular}{|l|lll|lll|}
\hline
\multirow{2}{*}{Adiabatically varying length  ($L_{h}=100nm$, $L_{c}=129nm$)} & \multicolumn{3}{c|}{QOR }    & \multicolumn{3}{c|}{QCR  }    \\ \cline{2-7} 
                           & \multicolumn{1}{l|}{|W|$_{max}$} & \multicolumn{1}{l|}{$COP_{max}$}& |W($COP_{max}$)| & \multicolumn{1}{l|}{|W|$_{max}$}& \multicolumn{1}{l|}{$COP_{max}$} & |W($COP_{max}$)| \\ \hline
                    $T_h = 2.49K$, $ 1<f<10$, $0<p<1$ & \multicolumn{1}{l|}{0.119}  & \multicolumn{1}{l|}{1.506}    & 0.119 &  \multicolumn{1}{l|}{0.119}  & \multicolumn{1}{l|}{1.500} & 0.119    \\ \hline

\end{tabular}}

\caption{Comparing magnitude of maximum work output and maximum COP of the system acting as quantum refrigerator during QOC and QCC. Work done here is in $\mu$eV.}
\label{Table4}
\end{table}

Tables \ref{Table3} and \ref{Table4} show that introducing an impurity has enhanced the work output of both QOC and QCC while keeping the efficiency constant. If we compute work for ISW without impurity (using Eq.~\ref{noimpeffo}) such that the length of ISW changes during the cycle($L_h = 100nm, L_c =163nm$), we get the work output of QOHE and QCHE as $27.2 \mu ev$ and $31.8 \mu ev$ respectively, which are lower than the work output produced by heat engine with impurity. The refrigerator phase of the thermodynamic cycle is absent for ISW without impurity. The maximum possible efficiency that can be achieved by any thermodynamic cycle is given by the Carnot efficiency limit, which is $1-\frac{T_{c}}{T_{h}}$.
From Table III and IV we notice that for QOC while varying length ($L_{h}$ = 100nm, $L_{c} $= 163nm) of ISW we could achieve efficiency near to the Carnot efficiency limit, given by $1-\frac{T_{c}}{T_{h}} = 0.7$.

Thus introducing an impurity in the ISW has unlocked the refrigerator phase in the thermodynamic cycle and has also led to the higher work output of the quantum heat engine than the quantum heat engine without impurity and has helped QOC achieve near Carnot efficiency.

\subsection{Adiabatically varying position of impurity during Otto cycle in weak coupling limit (\ref{IIIC})}
While adiabatically varying impurity position \emph{during} the cycle, we see both quantum heat engine and cold pump phases.


\begin{table}[H]
\hspace{-1.5cm}
\scalebox{0.72}{
\def\arraystretch{1.5} \small
\begin{tabular}{|l|lll|lll|}
\hline
\multirow{2}{*}{Adiabatically varying position ($p_{h}=0.1$, $p_{c}=0.8$)} & \multicolumn{3}{c|}{QOHE  }    & \multicolumn{3}{c|}{QOCP }    \\ \cline{2-7} 
                           & \multicolumn{1}{l|}{W$_{max}$} & \multicolumn{1}{l|}{$\eta_{max}$} & W($\eta_{max}$) & \multicolumn{1}{l|}{|W|$_{max}$} & \multicolumn{1}{l|}{$COP_{max}$} &|W($COP_{max}$)| \\ \hline
                    $T_h = 25K$, $ 20nm<L<80nm$, $1<f<6$ & \multicolumn{1}{l|}{7.5}  & \multicolumn{1}{l|}{0.035} & 5.0  & \multicolumn{1}{l|}{5.0}  &\multicolumn{1}{l|}{1750} & 0.5    \\ \hline
                    
             $L =40nm$, $5K<T_h<35K, 1<f<3$ & \multicolumn{1}{l|}{5.0}  & \multicolumn{1}{l|}{0.03} & 3.0    & \multicolumn{1}{l|}{10.0}  &\multicolumn{1}{l|}{800} & 1.0

          \\ \hline
                 $f= 5$, $5K<T_h<35K$, $20nm<L<80nm$ & \multicolumn{1}{l|}{4}  & \multicolumn{1}{l|}{0.008} & 2.0    & \multicolumn{1}{l|}{2.0}  &\multicolumn{1}{l|}{2000}  & 0.4   \\ \hline

\end{tabular}}

\caption{The table shows efficiency, coefficient of performance and maximum work output delivered by the system while acting as quantum heat engine and quantum cold pump during the QOC.Work done here is in $\mu$eV.}
\label{Table5}
\end{table}


In Fig.~\ref{Fig15}(a), we observe both the heat engine and cold pump phases. The phase changes from heat engine to cold pump when length of ISW crosses 50nm, provided the position of impurity changes during the cycle as $p_h = 0.1 $ and $p_c = 0.8$. In Fig.~\ref{Fig15}(b), we observe both the heat engine and cold pump phases. Keeping the length of ISW constant at $L=40nm$ and changing the temperature of the hot reservoir along with changing the impurity position during the cycle helps us see the phase change from heat engine to cold pump.
The maximum work output of heat engine and cold pump obtained while varying position of impurity adiabatically during QOC are tabulated in Table \ref{Table5}.

The cold pump phase of the thermodynamic cycle is absent when we use ISW without impurity. Introducing an impurity in the ISW has unlocked the cold pump phase in the thermodynamic cycle. In table VIII, we get large values of $COP_{max}$ due to large values of $Q_{out}$ and small values of work done in the cold pump phase.

\subsection{Adiabatically varying position of impurity in strong coupling limit (\ref{IIID})}
We can do a similar analysis in the strong coupling limit ($|f|\ll0.1$), just like the way we did for the weak coupling case ($|f|>0.5$). The energy eigenvalue for the strong coupling limit is derived in Eq.~(\ref{strongeigen}) up to first order in strength, i.e., $f$.

All three cases can be analyzed for this case, including the varying impurity position, length of ISW, and variable impurity strength during the cycle.

\begin{table}[H]
\hspace{-0.5cm}
\scalebox{0.7}{
\def\arraystretch{1.5} \small
\begin{tabular}{|l|lll|lll|}
\hline
\multirow{2}{*}{Adiabatically varying position ($p_{h}=0.2, p_{c}=0.5, 0.8$)} & \multicolumn{3}{c|}{QOHE }    & \multicolumn{3}{c|}{QOR  }    \\ \cline{2-7} 
                           & \multicolumn{1}{l|}{W$_{max}$} & \multicolumn{1}{l|}{$\eta_{max}$} & W($\eta_{max}$) &\multicolumn{1}{l|}{|W|$_{max}$} & \multicolumn{1}{l|}{COP$_{max}$} & |W(COP$_{max}$)|  \\ \hline
                    $T_h = 10K$, $ 0.01<f<0.08$, $20nm<L<80nm$ & \multicolumn{1}{l|}{0.003}  & \multicolumn{1}{l|}{0.847} & 0.001   & \multicolumn{1}{l|}{0.001}  &\multicolumn{1}{l|}{0.063} & 0.0002    \\ \hline
                    
             $f =0.03$, $5K<T_h<35K$, $20nm<L<80nm$ & \multicolumn{1}{l|}{0.1}  & \multicolumn{1}{l|}{0.94} & 0.01    & \multicolumn{1}{l|}{0.1}  &\multicolumn{1}{l|}{0.065} &0.03   \\    \hline

\end{tabular}}

\caption{Maximum work done on the system ($|W|_{max}$) and Work done at maximum COP, both in $meV$, with maximum COP, i.e., $COP_{max}$ when strength of impurity is varied adiabatically during Otto cycle.}
\label{Table6}
\end{table}

In strong coupling, we vary impurity position during the Otto cycle and study the density plots obtained. While adiabatically changing the position of impurity \emph{during} this strong coupling cycle, we see quantum heat engine and refrigerator phases. Interestingly, when impurity position varied during weak coupling, we could detect heat engine and cold pump phases only; however, when we varied position during strong coupling, we see refrigerator instead of cold pump phase. Table \ref{table8} compares the changes obtained in operational phases of the ISW after adding impurity with the no impurity case. 
The Carnot efficiency limit gives the maximum achievable efficiency of any cycle. From Table \ref{Table6}, we notice that while varying the position of the impurity in the strong coupling, we get the efficiency of QOC very near to the Carnot efficiency limit, which is $1-\frac{T_{c}}{T_{h}} = 0.85$ and $0.96$. Thus, the impurity has helped attain the maximum possible efficiency for the Otto cycle in strong coupling.

\onecolumngrid

\begin{table}[H]
\centering
\scalebox{1.2}{
\def\arraystretch{1.5} \small
\begin{tabular}{|c|l|l|l|l|c|}
\hline
\multicolumn{2}{|c|}{\multirow{2}{*}{ISW with and without impurity}} & \multicolumn{3}{c|}{With Impurity} & Without Impurity \\
\cline{3-6}
\multicolumn{2}{|c|}{} & \multicolumn{1}{c|}{Heat Engine} & Refrigerator & \multicolumn{1}{c|}{Cold Pump} & \multicolumn{1}{l|}{Heat Engine} \\
\hline
\multirow{3}{*}{QOC} & Changing strength of impurity & Present & Absent & Present & Absent \\
\cline{2-6}
& Changing length of ISW & Present & Present & Absent & Present \\
\cline{2-6}
& Changing position of impurity & Present & Absent & Present & Absent \\
\hline
\multirow{3}{*}{QCC} & Changing strength of impurity & Absent & Absent & Absent & Absent \\
\cline{2-6}
& Changing length of ISW & Present & Present & Absent & Present \\
\cline{2-6}
& Changing position of impurity & Absent & Absent & Absent & Absent \\
\hline
\end{tabular}}
\caption{Heat engine, refrigerator and cold pump phases in an ISW with and without impurity}
\label{table8}
\end{table}
\twocolumngrid
\section{Conclusions and Experimental realization}
Adding an impurity to an ISW as the thermodynamic system can unveil different thermodynamic phases like a quantum heat engine, refrigerator, and cold pump. Introducing the impurity has also resulted in higher work outputs for QOHE and QCHE. We present the analytical perturbative eigenenergy correction up to second order for an ISW with impurity in weak coupling and up to first order with impurity in a strong coupling regime. We show that the weak coupling perturbative solution is in good agreement with the numerical solution of the transcendental dispersion relation. We find the threshold value for the impurity strength for which weak coupling perturbative results can be applied. We analyze this system by varying the position of impurity, length of ISW, temperature of reservoirs, and strength of the impurity. We have considered all possible system variants while tuning all possible system parameters. When the length of the ISW is varied during the cycle, the Carnot reversibility condition gets satisfied. Hence, we could compare work and efficiency for both QCC and QOC. QCHE can generate a higher work output and efficiency than QOHE. Further, we could also unlock the quantum refrigerator phase in QCC and QOC. 
\\
A possible candidate to realize the model experimentally, involves a laser-cooled trapped ion as a microscopic heat machine Ref.~\cite{Rajibul}. In Ref.~\cite{Rajibul}, not only is the potential experimentally realized which mimics our model (ISW with impurity), but also construction and demonstration of quantum Otto engine is carried out experimentally. A trapped, laser-cooled ion with the combined electro-static
harmonic potential of a Paul ion trap and a sinusoidal potential of an optical lattice can be used to mimic an infinite square well both with and without impurity. Further, we list a few experimental candidates to realize the ISW with/without impurity using quantum dots. In Ref.~\cite{Hydrogenic_impurities}, the authors work with hydrogenic impurities in
GaAs-(Ga,Al)As quantum dots to create a finite confining spherical potential well
with depth determined by the discontinuity of the band gap in the quantum dot. Calculations were also performed for an infinite spherical confining potential, similarly, in  Ref.~\cite{Feng}
a quantum wire is realized into which an impurity of variable size is introduced, which can be used to create quantum dots. The quantum wells created in such systems can also be made to interact experimentally, for example, as shown in Ref.~\cite{LDHICKS} for multiple quantum wells grown using PbTe/Pb$_{1-x}$Eu$_{x}$Te in a molecular beam epitaxy. These low dimensional structures depict quantum dot superlattices leading to strongly quantized energy spectrum of electron. The super lattice quantum well structure relies on high tunnelling probability of the electron. The electron is no longer localized inside an individual quantum well, thus the wells can be made interacting. A way to harness heat flow in such systems
is shown in Ref.~\cite{Liu} where a heat engine composed of serially connected two quantum dots sandwiched between two metallic electrodes is proposed. These works can be extended by trying out different types of working substances in order to unlock the exotic properties of quantum heat cycles, like in Ref.~\cite{PhysRevE.104.014149}, a conceptual
design for quantum heat machines using a pair of coupled double quantum
dots is presented, with each pair containing as an excess electron, as the working substance.  
All the afore-mentioned works depict experimental techniques for creating systems quite close to our model consisting of an ISW with impurity. To summarize we say that presence of an impurity in a system with ISW potential can open new operational phases in the thermodynamic cycle and can also enhance the work output and the cycle's efficiency. 


\acknowledgments{CB acknowledges support from the  grant ``Josephson junctions with strained Dirac materials and their applicationin quantum information processing'' of Science \& Engineering Research Board (SERB) DST, Govt. of India, Grant No. CRG/20l9/006258.}
\bibliographystyle{ieeetr}
\bibliography{bibfile}
\end{document}